\documentclass[3p,12pt]{elsarticle}
\usepackage{booktabs}
\usepackage[fleqn]{amsmath}
\usepackage{cases}
\usepackage{multirow}
\usepackage{xcolor}
\usepackage[normalem]{ulem}
\usepackage{tabularx}
\usepackage{siunitx}
\usepackage{comment}
\usepackage{algorithm}
\usepackage{algpseudocode}
\usepackage{algorithmicx}
\usepackage{grffile}
\usepackage{rotating}
\usepackage{lineno}
\usepackage{hyperref}
\usepackage{graphicx,enumerate,subfigure}
\usepackage{amssymb}
\usepackage{bm,amsmath}
\usepackage{caption,color,xcolor}
\usepackage{subeqnarray}
\usepackage{multirow}
\usepackage{lscape}
\usepackage{rotating}
\usepackage{graphics}
\usepackage{float}
\usepackage{epstopdf,epsfig}
\usepackage{threeparttable}
\usepackage{soul}
\usepackage{enumitem}

\usepackage[capitalize]{cleveref}

\floatname{algorithm}{Algorithm}
\biboptions{numbers,sort&compress} 
\journal{XXX}

\newcounter{bla}

\usepackage{CJKutf8}

\begin{document}
\begin{CJK}{UTF8}{gbsn}



\begin{frontmatter}

\title{Frequency-domain general synthetic iterative scheme for efficient simulation of  oscillatory rarefied gas flows}

\author{Pengshuo Li} 
\author{Lei Wu\corref{mycorrespondingauthor}}
\cortext[mycorrespondingauthor]{Corresponding author}
\ead{wul@sustech.edu.cn}
 
\address{Department of Mechanics and Aerospace Engineering, Southern University of Science and Technology, Shenzhen 518055, China}

\begin{abstract}
Oscillatory rarefied gas flows are frequently encountered in micro-electro-mechanical systems, and their efficient numerical simulation remains a major challenge due to the time-dependent nature of the problem and the high dimensionality of the Boltzmann kinetic equation. Here, we address this challenge by focusing on the periodic steady state and solving the resulting problem using the frequency-domain general synthetic iterative scheme (GSIS). The key idea of GSIS is to simultaneously solve the mesoscopic kinetic equation and the macroscopic synthetic equation. The kinetic equation provides high-order constitutive relations—-beyond those given by Newton’s law of viscosity and Fourier’s law of heat conduction—-to the synthetic equation. In turn, the synthetic equation, which converges to the periodic steady state much faster than the kinetic equation, boosts the evolution of the kinetic equation toward the periodic steady state. As a result, super convergence is achieved, together with an asymptotic-preserving  property that allows the use of coarse spatial grids. The analytical Fourier stability analysis and the Chapman–Enskog expansion, together with challenging numerical simulations, are employed to demonstrate the fast convergence and asymptotic-preserving properties of GSIS, revealing that it can be three orders of magnitude faster than conventional kinetic schemes in near-continuum flow regimes.
\end{abstract}


\end{frontmatter}

\section{Introduction}

Micro-electro-mechanical systems (MEMS), such as microcantilever sensors and accelerometers, are widely used in sensing, actuation, and signal-processing applications \cite{Li2008BottomUp}. Their dynamic performance is strongly affected by gas damping induced by the interaction between the device and the surrounding gas \cite{Karniadakis2005Microflows}. 
Early analyses of gas damping were largely built on continuum descriptions. For example, squeeze-film damping (SFD) is often modeled using lubrication theory and the Reynolds equation (with the mass flow rate calculated from the Navier-Stokes-Fourier (NSF) equations) \citep{bao2007squeeze}, while oscillatory shear damping is commonly described by unsteady boundary-layer analyses~\citep{schlichting2016boundary}. 

As MEMS devices continue to shrink in size and operate at increasingly high frequencies, gas flows often enter regimes where rarefaction and strong unsteadiness coexist. Under such conditions,  the continuum assumptions underlying the NSF equations may break down when either the spatial Knudsen number (defined as the ratio of the molecular mean free path to the characteristic flow length) or the temporal Knudsen number (defined as the oscillatory frequency to the molecular mean collision frequency) is large \cite{LeiBook}, and kinetic descriptions based on the Boltzmann equation become necessary to accurately capture gas dynamics across a wide range of length and time scales \citep{chapman1990mathematical}.

The Boltzmann equation can be solved using both stochastic and deterministic approaches, such as the direct simulation Monte Carlo (DSMC) method~\citep{Bird1994DSMC} and discrete velocity method~\citep{Chu1965a}. In typical MEMS applications, the flow velocity is orders of magnitude smaller than the molecular thermal speed, which leads to severe statistical fluctuations in conventional DSMC simulations~\citep{hadjiconstantinou2003statistical,rader2011dsmc}. To address this issue, the low-variance DSMC~\citep{baker2005variance} has been developed. However, it requires stringent spatial and temporal resolutions and extensive sampling, so the computational efficiency deteriorates rapidly in near-continuum regimes. 
An alternative approach is to formulate oscillatory gas damping problems in the frequency domain, which yields the complex periodic response without long transient simulations~\citep{LadigesSader2015JCP,ladiges2015frequency}. 
However, frequency-domain DSMC does not asymptotically preserve (AP) the NSF limit; it still requires molecular-scale resolution in near-continuum regimes.

The discrete velocity method offers a noise-free alternative to particle simulations and is attractive for small-amplitude oscillatory flows, particularly when a frequency-domain kinetic formulation is used~\cite{wu2014oscillatory}. 
However, most discrete velocity methods rely on the conventional iterative scheme (CIS) with a split transport-collision treatment, which is not AP toward the NSF limit and whose convergence deteriorates sharply in near-continuum regimes, leading to mesh-sensitive accuracy and slow convergence~\cite{wang2018cf}. 
Unified gas-kinetic scheme~\cite{xu2010unified} and its discrete version~\cite{guo2013discrete} remedy this multiscale deficiency by coupling transport and collision in the flux evaluation, thereby achieving AP property and relaxing the mean-free-path constraint on the mesh~\citep{liu2014ugks,wang2018cf,zhang2021oscillatory}. 
For oscillatory rarefied gas flows, the unified gas-kinetic scheme is usually implemented in the time domain, so the amplitude and phase must be obtained from long-time integration after transients decay, which can be expensive in near-continuum regimes~\citep{wang2018nonlinear}.

Very recently, the general synthetic iterative scheme (GSIS) has emerged as a multiscale method for solving the Boltzmann kinetic equation with fast-converging and AP properties~\citep{su2020can, su2020fast}. Specifically, the steady-state solution can be obtained within dozens of iterations, and the spatial cell size can be much larger than the mean free path. Motivated by the need for accurate and efficient simulation of oscillatory rarefied gas flows, we develop a GSIS for the frequency-domain linearized Boltzmann equation. This approach combines AP properties with super-convergence, enabling efficient simulation of linear oscillatory gas flows over a wide range of Knudsen numbers and oscillation frequencies.

The remainder of this paper is organized as follows. 
Sections~\ref{sec:CIS} and~\ref{sec:GSIS} present the formulations of CIS and GSIS, respectively.
The super-convergence and AP properties of the frequency-domain GSIS are then validated via a Fourier stability analysis and the Chapman-Enskog expansion.
Section~\ref{sec:numerical} presents the numerical methodology and implementation details. Section~\ref{sec:result} discusses representative examples, including oscillatory shear-driven flows and SFD problems, to demonstrate the accuracy and efficiency of the proposed method. 
Conclusions are drawn in Section~\ref{sec:conclusion}.


\section{Gas kinetic equation and CIS}
\label{sec:CIS}
In this section, we present the gas kinetic equation and analyze the convergence rate of CIS for finding the periodic steady state.

\subsection{Kinetic equation}
We investigate the response of the gas to a harmonically oscillating wall, whose velocity is prescribed as $ U_w(t)=\Re\!\left\{U_0 e^{\imath\omega t}\right\}$, where $U_0$ and $\omega$ are respectively the amplitude and angular frequency of vibration, $\imath$ is the imaginary unit, $t$ is the time, and $\Re\{\cdot\}$ denotes the real part of a complex number. Assume $U_0$ is much smaller than the most probable molecular speed $v_m$, that is, 
\begin{equation}\label{xi_def}
   \xi=\left|\frac{U_0}{v_m}\right|\ll1, \quad \text{with}~ v_m= \sqrt{\frac{2k_B T_0}{m}},
\end{equation}
where $k_B$ is the Boltzmann constant and $m$ is the molecular mass of the gas. In this case, the velocity distribution function for the gas molecules can be written as
\begin{equation}
    f(\bm{v},\bm{x},t) = \Re
    \left\{f_{eq}(\bm{v}) \left[ 1 + \xi h(\bm{{v}},\bm{x})\exp(\imath\omega t) \right]\right\},
\end{equation}
where $\bm{v}=({v}_x,{v}_y,{v}_z)$ is the molecular velocity, $\bm{x}=(x,y,z)$ is the spatial coordinate vector, and $h$ denotes the perturbation from the global equilibrium state $f_{eq}(\bm{v})$.

When the Shakhov model~\cite{shakhov1968approximate} is used, the frequency-domain linearized Boltzmann equation can be written as
\begin{equation}\label{eq:dvmeqn}
\begin{gathered}
    \imath{S} h + \bm{v} \cdot \nabla h = \delta_{rp} \left[\rho + 2\bm{v} \cdot \bm{u}
    + \left(v^2 - \frac{3}{2}\right)\tau
    + \frac{4}{15}\left(v^2 - \frac{5}{2}\right)\bm{v} \cdot \bm{q} - h \right],
\end{gathered}
\end{equation}
where 
\begin{equation}
    S = \frac{\omega L}{v_m}
\end{equation}
is the Strouhal number, and
\begin{equation}
    \delta_{rp} = \frac{p_0 L}{\mu v_m} 
\end{equation}
is the rarefaction parameter 
($L$ is the reference length, $p_0$ is the reference pressure, and $\mu$ is the dynamic viscosity of the gas at the reference temperature $T_0$). Note that the rarefaction parameter is related to the spatial Knudsen number $Kn$ as $\delta_{rp}=\sqrt{\pi}/(2Kn)$. The temporal Knudsen number $Kn_t$ can be expressed as 
\begin{equation}\label{temporal_Kn}
    Kn_t=\delta^{-1}_{rp} S.
\end{equation}

Note that the distribution function is normalized by \(n_0/v_m^3\) (\(n_0\) is the reference number density), 
the spatial coordinates by \(L\), the molecular velocity by \(v_m\), and time by \(L/v_m\).
The corresponding macroscopic moments are
\begin{equation}\label{eq:moment}
    \begin{bmatrix}
    \rho \\ 
    {\bm{u}} \\ 
    \tau \\ 
    \bm{\Pi} \\ 
  {\bm{q}}
    \end{bmatrix}
    =
    \int
    \begin{bmatrix}
    1 \\
    \bm{v} \\
    \frac{2}{3}v^2-1 \\ 
    2\left(\bm{v}\bm{v}-\frac{v^2}{3}\bm{I}\right) \\
    \left(v^2-\frac{5}{2}\right)\bm{v}
    \end{bmatrix}
    f_{eq} h \, d\bm{v},
\end{equation}
where $\rho$ and $\tau$ represent the deviations from the reference density and temperature, which are normalized by $n_0$ and $T_0$, respectively; $\bm{\Pi}$ and ${\bm{q}}$ denote the deviatoric stress and heat flux, which are normalized by $n_0k_BT_0$ and $n_0k_BT_0v_m$, respectively. $\bm{I}$ is the identity matrix. The global equilibrium state $f_{eq}(\bm{v})$ is
\begin{equation}
    f_{eq}(\bm{v}) = \pi^{-3/2}\exp(-v^2).
\end{equation}
Besides these standard normalizations, all macroscopic quantities in Eq.~\eqref{eq:moment} are further normalized by the relative magnitude of perturbation $\xi$ in Eq.~\eqref{xi_def}.

\subsection{CIS and its convergence rate}

In CIS, given the value of velocity distribution function $h^n$ at the $n$-th iteration, the value at the next iteration $h^{n+1}$ is obtained by solving the following equation:
\begin{equation}\label{eq:CIS}
    \begin{gathered}
        (\imath{S}+\delta_{rp}) h^{n+1} + \bm{v} \cdot \nabla h^{n+1} = \delta_{rp} \left[\rho^n + 2\bm{v} \cdot \bm{u}^n
        + \left(v^2 - \frac{3}{2}\right)\tau^n
        + \frac{4}{15}\left(v^2 - \frac{5}{2}\right)\bm{v} \cdot \bm{q}^n \right].
    \end{gathered}
\end{equation}

For simplicity, we present the stability analysis for two-dimensional perturbations. First, we define the error function between distribution functions at two consecutive iterations as
\begin{equation}\label{eq:Y}
    Y^{n+1}(\bm{x},\bm{v})=h^{n+1}(\bm{x},\bm{v})-h^n(\bm{x},\bm{v}),
\end{equation}
and the error function between macroscopic quantities $M=[\rho,\bm{u},\tau,\bm{q}]^\top$ (\(\top\) denotes the transpose operator) as
\begin{equation}
    \begin{aligned}
        \bm{\Phi}^{n+1}(\bm{x})=&\bm{M}^{n+1}(\bm{x})-\bm{M}^n(\bm{x}) \\
        =&\int Y^{n+1}(\bm{x},\bm{v})\bm{\phi}(\bm{v}) \, d\bm{v},
    \end{aligned}
    \end{equation}
where
\begin{equation}\label{eq:phi}
    \bm{\phi}(\bm{v})=
    \left[1,
    v_x,v_y,
    \frac{2}{3}v^2-1,
    v_x\left(v^2-\frac{5}{2}\right),
    v_y\left(v^2-\frac{5}{2}\right)\right]^\top.
\end{equation}
Second, to determine the error convergence rate $e$, we express the error functions as
\begin{equation}
    \begin{aligned}\label{eq:Y,phi}
       & Y^{n+1}(\bm{x},\bm{v})=e^n\bar{Y}(\bm{v})\exp(\imath \bm{\theta} \cdot \bm{x}),\\
      &  \bm{\Phi}^{n+1}(\bm{x})=e^{n+1}
        \bm{\alpha}
        \exp(\imath\bm{\theta} \cdot \bm{x}),
    \end{aligned}
\end{equation}
where $\bm{\theta}=(\theta_x,\theta_y)$ is the perturbation wave vector, and $\bm{\alpha}$ is the amplitude vector of the macroscopic error mode,
\(\bm{\alpha}=[\alpha_\rho,\alpha_{u_x},\alpha_{u_y},\alpha_\tau,\alpha_{q_x},\alpha_{q_y}]^\top\).
Without loss of generality, we set \(|\bm{\theta}|=1\) to focus on the dependence on the spatial Knudsen number and Strouhal number.

From Eqs.~\eqref{eq:phi} and~\eqref{eq:Y,phi}, we have
\begin{equation}\label{eq14}
    e\bm{\alpha} = \int\bar{Y}(\bm{v})\bm{\phi}(\bm{v})d\bm{v}.
\end{equation}
where the expression for $\bar{Y}$ can be derived from Eqs.~\eqref{eq:dvmeqn},~\eqref{eq:Y}, and~\eqref{eq:Y,phi}  as
\begin{equation}\label{eq:Y(c)}
    \bar{Y}(\bm{v})=\frac{\alpha_\rho
    +2\alpha_{\bm{u}} \cdot \bm{v}
    +\alpha_\tau\left(v^2-\frac{3}{2}\right)
    +\frac{4}{15} (\alpha_{\bm{q}} \cdot \bm{v}) \left(v^2-\frac{5}{2}\right)}
    {1+\imath\delta^{-1}_{rp}(S+\bm{\theta} \cdot \bm{v})}
    f_{eq}.
\end{equation}

Finally, substituting Eq.~\eqref{eq:Y(c)} into Eq.~\eqref{eq14}, we obtain the linear system $C \bm{\alpha}=e\bm{\alpha}$, where $C$ is a $6\times6$ matrix. The error decay rate is characterized by the spectral radius \(\rho(C)=\max_i|\lambda_i|\), where \(\lambda_i\) are the eigenvalues of \(C\). 
As shown in Fig.~\ref{fig:converge_CIS_GSIS}, when the Knudsen number is large, $e$ tends to 0, indicating that the convergence speed of CIS is very fast. However, when the Knudsen number is small, i.e., the flow approaches the hydrodynamic regime, $e$ approaches 1, indicating that the change between adjacent iterations is very small, and thus the convergence speed becomes very slow. 

Moreover, the CIS exhibits the phenomenon of false convergence~\cite{ADAMS2023}. Specifically, when $S$ is small, the error decay rate can be approximated by $1 - O(1/\delta_{rp}^2)$ in the near-continuum flow regime. If the absolute difference between two consecutive iterations is $\epsilon$, then the absolute difference between the final iteration and the true solution is on the order of \cite{wu2022rarefied}
\begin{equation} \label{CIS_false}
   \frac{\rho(\bm{C})}{1-\rho(\bm{C})} \epsilon 
   \rightarrow \delta_{rp}^2 \epsilon, ~\text{when}~ \delta_{rp}\rightarrow\infty.
\end{equation}
Consequently, even when $\epsilon$ is small, the overall error remains large.


\section{General synthetic iterative scheme (GSIS)}
\label{sec:GSIS}
To tackle the slow convergence in near-continuum flows, GSIS employs macroscopic synthetic equations to accelerate the global error decay. 
First, when the velocity distribution function and the corresponding macroscopic quantities are known at the $n$-th step, the intermediate distribution function  $h^{n+\frac{1}{2}}$ is obtained by performing one CIS update:
\begin{equation}\label{eq:GSIS_h}
    (\imath{S}+\delta_{rp}) h^{n+\frac{1}{2}} + \bm{v} \cdot \nabla h^{n+\frac{1}{2}} = \delta_{rp} \left[\rho^n + 2\bm{v} \cdot \bm{u}^n
    + \left(v^2 - \frac{3}{2}\right)\tau^n
    + \frac{4}{15}\left(v^2 - \frac{5}{2}\right)\bm{v} \cdot \bm{q}^n  \right].
\end{equation}

Second, the kinetic equation~\eqref{eq:dvmeqn} is multiplied by $[1, \bm{v}, \frac{2}{3}v^2-1]f_{eq}$ and integrated over the velocity space, yielding the evolution equations for the density, velocity, and temperature:
\begin{equation}	\label{eq:sysEqn}
\begin{aligned}
&	\imath{} S \rho + \nabla \cdot \bm{u} = 0, \\ 
& 2\imath{} S \bm{u} + \nabla\rho + \nabla\tau + \nabla \cdot \bm{\Pi} = 0, \\ 
&	\frac{3}{2}\imath{} S \tau + \nabla \cdot \bm{u} + \nabla \cdot \bm{q} = 0.
\end{aligned}
\end{equation}

Obviously, Eq.~\eqref{eq:sysEqn} is not closed, as the stress tensor $\bm{\Pi}$ and the heat flux $\bm{q}$ remain unknown. 
The Chapman-Enskog expansion~\cite{chapman1990mathematical} of the kinetic equation to the first order of Knudsen number (the NSF approximation) gives the linear constitutive relations as
\begin{equation}    	\label{eq:NSF_consitu}
	\begin{aligned}
	\bm{\Pi}_\text{NSF} = -\frac{1}{\delta_{rp}}\left[\nabla \bm{u} + (\nabla \bm{u})^\top - \frac{2}{3}(\nabla \cdot \bm{u})\bm{I}\right], 
    \quad
	\bm{q}_\text{NSF} = -\frac{15}{8\delta_{rp}}\nabla{\tau}.
	\end{aligned}
\end{equation}
However, the linear NSF constitutive relations are no longer accurate in highly rarefied regimes.


\subsection{Original GSIS}
For original GSIS, the evolution equations for stress and heat flux are derived by further multiplying the kinetic equation~\eqref{eq:dvmeqn} by $[2(\bm{v}\bm{v}-\frac{v^2}{3}\bm{I}), (v^2-\frac{5}{2})\bm{v}]f_{eq}$ and integrating over the velocity space, yielding
\begin{equation}	\label{eq:stress_heat_evolve}
	\begin{aligned}
	&	\imath{S} \bm{\Pi} + 
		\int 2\left(\bm{v}\bm{v}-\frac{v^2}{3}\bm{I}\right)\,\bm{v}\cdot\nabla h f_{eq}d\bm{v}
		= -\delta_{rp}\bm{\Pi}, \\
	&	\imath{S} \bm{q} +
		\int \left(v^2-\frac{5}{2}\right)\bm{v}(\bm{v}\cdot\nabla h) f_{eq}d\bm{v}
		= -\frac{2}{3}\delta_{rp}\bm{q}.
	\end{aligned}
\end{equation}

Following Ref.~\cite{su2020can}, Eq.~\eqref{eq:stress_heat_evolve} can be rearranged as
\begin{equation}\label{eq:stress_heat_evolve0}
	\begin{aligned}
	&	\imath{S} \bm{\Pi} + 
		\int 2\left(\bm{v}\bm{v}-\frac{v^2}{3}\bm{I}\right)\,\bm{v}\cdot\nabla h^{n+\frac{1}{2}}f_{eq}d\bm{v}
		= -\delta_{rp}\left(\bm{\Pi} -\bm{\Pi}_\text{NSF}+\bm{\Pi}^{n+\frac{1}{2}}_\text{NSF} \right), \\
	&	\imath{S} \bm{q} +
		\int \left(v^2-\frac{5}{2}\right)\bm{v}(\bm{v}\cdot\nabla h^{n+\frac{1}{2}}) f_{eq}d\bm{v}
		= -\frac{2}{3}\delta_{rp}
        \left( \bm{q}-\bm{q}_\text{NSF} +\bm{q}^{n+\frac{1}{2}}_\text{NSF} \right),
	\end{aligned}
\end{equation}
where quantities with superscript \(n+\tfrac12\) are evaluated at the intermediate step, whereas the unsuperscripted  are evaluated at the \((n+1)\)-th iteration.
Accordingly, \(\bm{\Pi}\) and \(\bm{q}\) used to close Eq.~\eqref{eq:sysEqn} are evaluated as
\begin{equation} \label{eq:GSISI-stress_heat}
    \begin{aligned}
    &    \bm{\Pi} = \frac{1}{\imath{S} + \delta_{rp}}
        \left[\delta_{rp}\bm{\Pi}_\text{NSF} - \delta_{rp}\bm{\Pi}^{n+\frac{1}{2}}_\text{NSF}- \int 2\left(\bm{v}\bm{v}-\frac{v^2}{3}\bm{I}\right)\,\bm{v}\cdot\nabla h^{n+\frac{1}{2}} f_{eq}d\bm{v}
	 \right], \\
    &    \bm{q} = \frac{1}{\imath{S} + 
        \frac{2}{3}\delta_{rp}}
        \left[\frac{2}{3}\delta_{rp}\bm{q}_\text{NSF} 
        -\frac{2}{3}\delta_{rp}\bm{q}^{n+\frac{1}{2}}_\text{NSF}
        -\int \left(v^2-\frac{5}{2}\right)\bm{v}\left(\bm{v}\cdot\nabla h^{n+\frac{1}{2}}
        \right) f_{eq}d\bm{v}
        \right].
    \end{aligned}
\end{equation}

To calculate the convergence rate of original GSIS, we redefine the error functions as
\begin{equation}
    \begin{aligned}\label{eq:Y,phi,GSIS}
     &   Y^{n+\frac{1}{2}}(\bm{x},\bm{v})=
     h^{n+\frac{1}{2}}(\bm{x},\bm{v})
     -h^n(\bm{x},\bm{v})
     =e^n\bar{Y}(\bm{v})\exp(\imath\bm{\theta}\cdot \bm{x}), \\
    &    \Phi^{n+1}=
    M^{n+1}(\bm{x})
    -M^n(\bm{x})=
    e^{n+1}\bm{\alpha}\exp(\imath\bm{\theta}\cdot \bm{x}),
    \end{aligned}
\end{equation}
where \(\bar{Y}(\bm{v})\) is given by Eq.~\eqref{eq:Y(c)}, while the macroscopic variables at the \((n+1)\)-th step are obtained by solving the synthetic equations.
Using Eqs.~\eqref{eq:Y,phi,GSIS},~\eqref{eq:sysEqn} and~\eqref{eq:stress_heat_evolve}, we obtain a linear system of equations for $\alpha$:
\begin{equation}\label{eq:GSISI-Matrix}
    \begin{aligned}
        &e\left[\imath{S}\alpha_\rho+\imath\bm{\theta} \cdot \alpha_{\bm{u}}\right]=0, \\
        &e\left[\imath\theta_x(\alpha_\rho+\alpha_\tau)
        +\left(2\imath{S}+\frac{1}{\imath{S}+\delta_{rp}}\right)\alpha_{u_x} \right]=S_2, \\
        &e\left[\imath\theta_y(\alpha_\rho+\alpha_\tau)+
        \left(2\imath{S}+\frac{1}{\imath{S}+\delta_{rp}}\right)\alpha_{u_y}\right]=S_3, \\
        &e\left[\frac{3}{2}\imath{S}\alpha_\tau
        +\imath \bm{\theta} \cdot \alpha_{\bm{u}} 
        +\imath \bm{\theta} \cdot \alpha_{\bm{q}}\right]=0, \\
        &e\left[\frac{15}{8}\imath\theta_x\delta^{-1}_{rp} \alpha_\tau 
        +\left(1+\frac{3}{2}\imath{S}\delta^{-1}_{rp}\right)\alpha_{q_x}\right]=S_5,   \\
        &e\left[\frac{15}{8}\imath\theta_y \delta^{-1}_{rp} \alpha_\tau
        +\left(1+\frac{3}{2}\imath{S}\delta^{-1}_{rp}\right)\alpha_{q_y}\right]=S_6,      
        \end{aligned}
\end{equation}
where the source terms, due to the quantities at the intermediate step in Eq.~\eqref{eq:GSISI-stress_heat}, are expressed as
\begin{equation}\label{eq:GSISI-source}
    \begin{aligned}
        &S_2=\frac{1}{\imath{S}+\delta_{rp}}\int \left[
        v_x-2\bm{\theta} \cdot \bm{v} \left(
        \theta_x \left(v_x^2-\frac{v^2}{3}\right)+\theta_y v_x v_y
        \right) \right]
        \bar{Y}(\bm{v})\,d\bm{v}, \\
        &S_3=\frac{1}{\imath{S}+\delta_{rp}}\int \left[
        v_y-2\bm{\theta} \cdot \bm{v} \left(\theta_x v_x v_y+\theta_y\left(v_y^2 -\frac{v^2}{3}\right)\right)
        \right] 
        \bar{Y}(\bm{v})d\bm{v}, \\
        &S_5=\frac{\imath}{\delta_{rp}}\int\left[
        \frac{15}{8}\theta_x\left(\frac{2}{3}v^2-1\right)
        -\frac{3}{2}\bm{\theta} \cdot \bm{v} v_x\left(v^2-\frac{5}{2}\right)\right]
        \bar{Y}(\bm{v})d\bm{v},  \\
        &S_6=\frac{\imath}{\delta_{rp}}\int\left[
        \frac{15}{8}\theta_y\left(\frac{2}{3}v^2-1\right)
        -\frac{3}{2} \bm{\theta} \cdot \bm{v} v_y\left(v^2-\frac{5}{2}\right) \right]
        \bar{Y}(\bm{v})d\bm{v}. 
    \end{aligned}
\end{equation}

\begin{figure}[t]
    \centering
    \subfigure[\(Kn=10\)]{\includegraphics[width=0.4\linewidth]{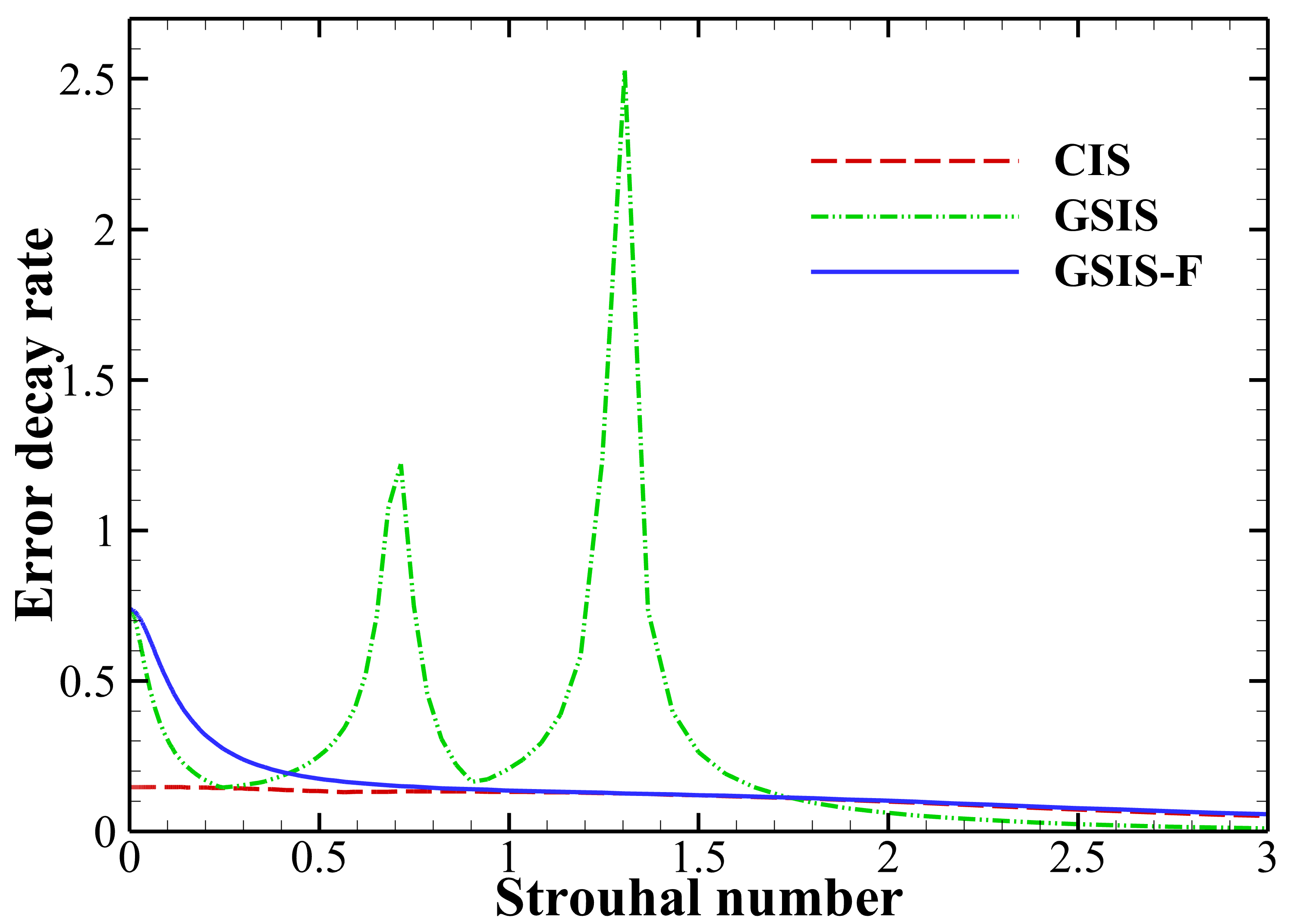}}
    \subfigure[\(Kn=1\)]{\includegraphics[width=0.4\linewidth]{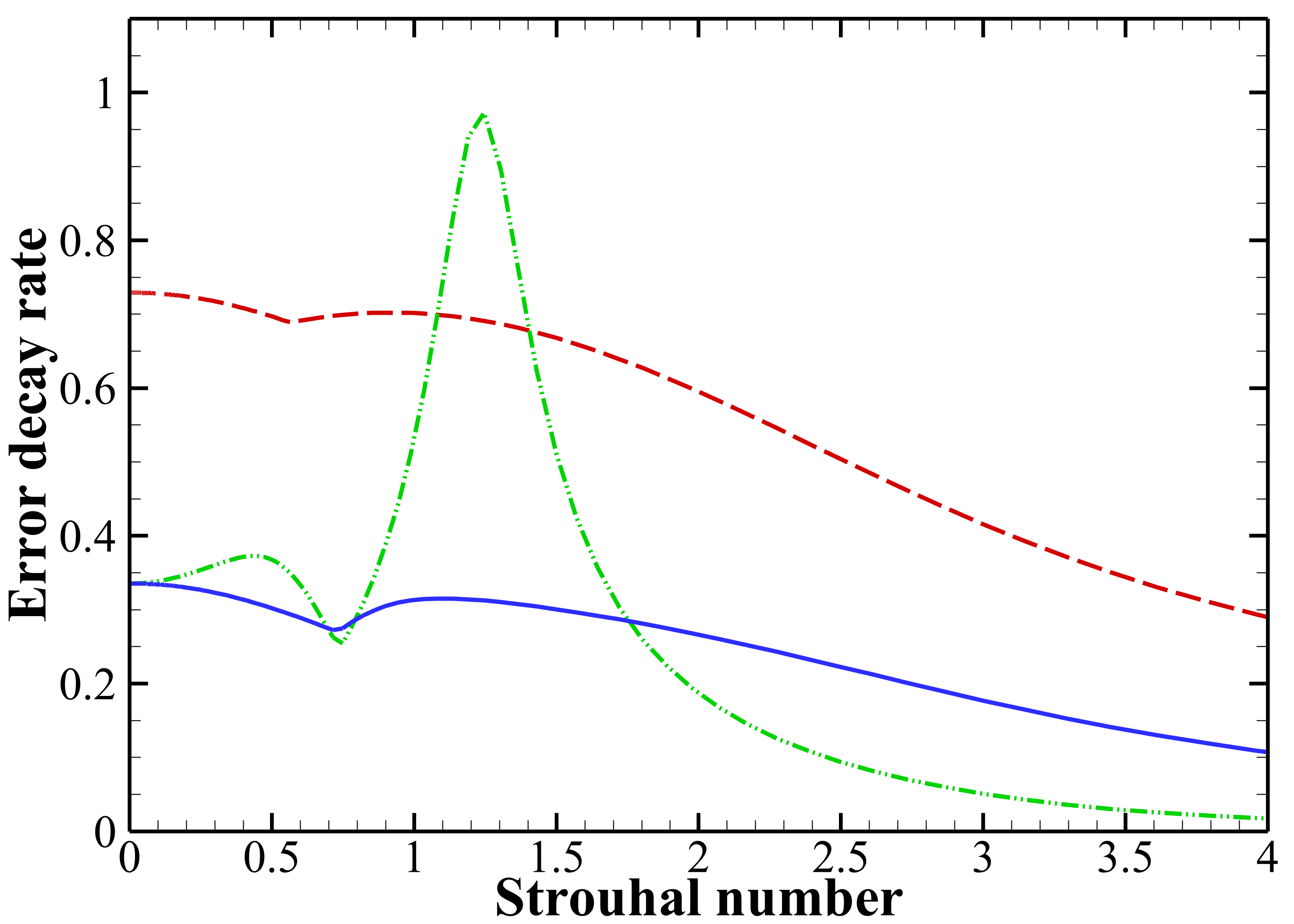}}\\
    \subfigure[\(Kn=0.1\)]{\includegraphics[width=0.4\linewidth]{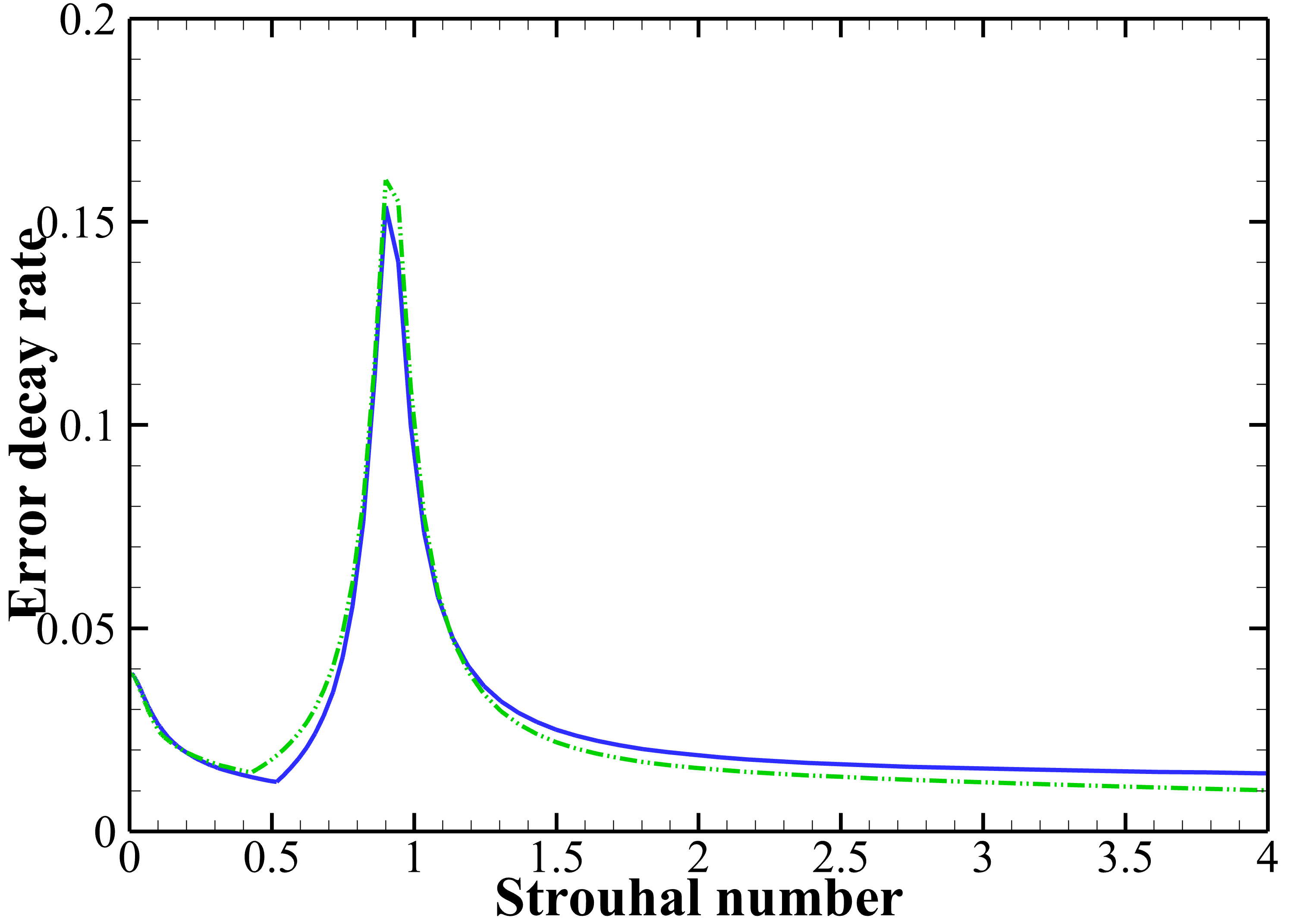}}
    \subfigure[\(Kn=0.01\)]{\includegraphics[width=0.4\linewidth]{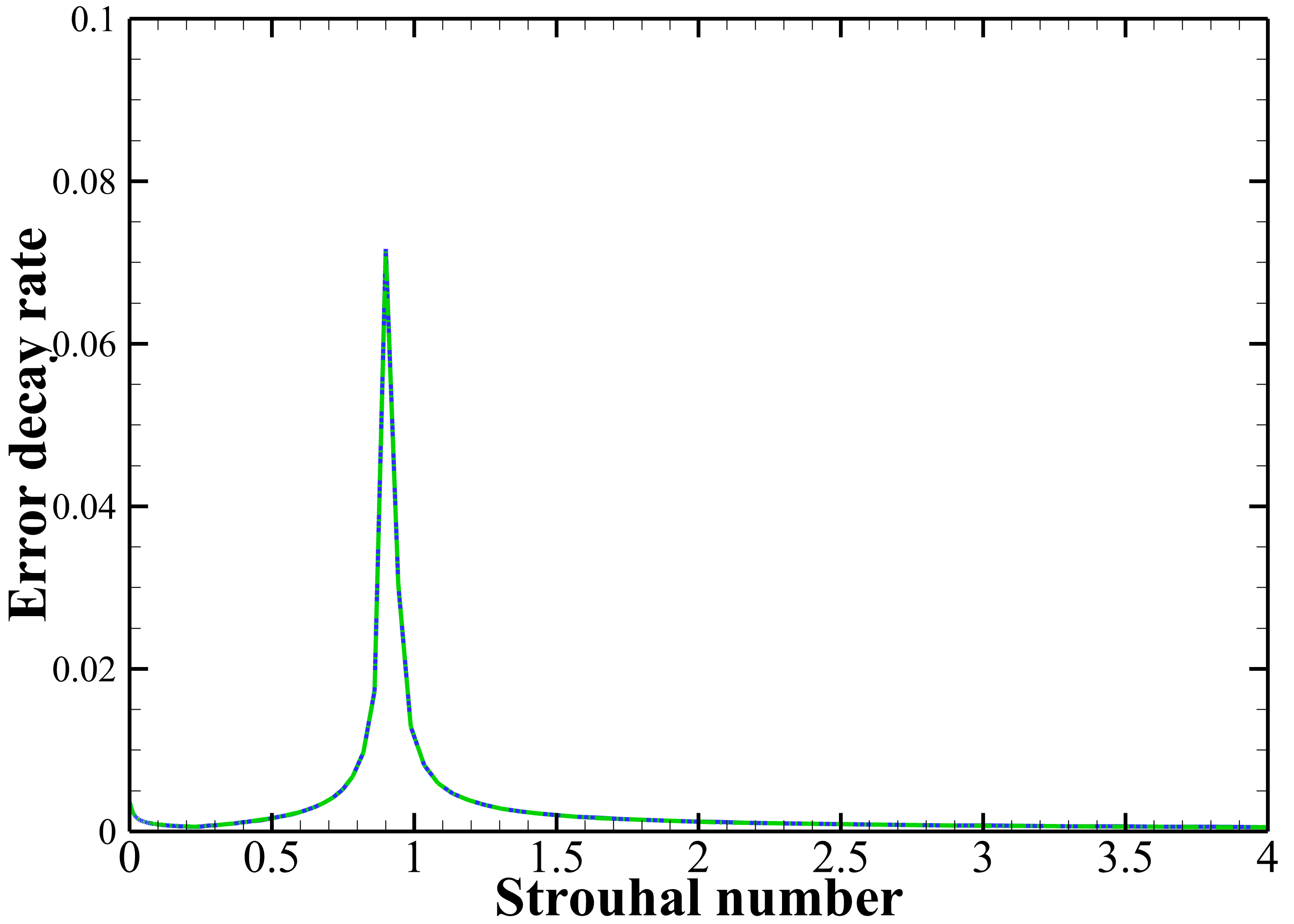}}
    \caption{Error decay rate of CIS and GSIS variants as functions of the Strouhal number \(S\) for different Knudsen numbers \(Kn\). For brevity, the original GSIS is denoted by GSIS and the frequency-domain GSIS by GSIS-F. In (c) and (d), the CIS decay rate is very close to unity for \(S<4\);  therefore, it is omitted for clarity and only GSIS and GSIS-F are shown.}
    \label{fig:converge_CIS_GSIS}
\end{figure}

The error convergence rate of original GSIS can be obtained by solving Eqs.~\eqref{eq:GSISI-Matrix} and~\eqref{eq:GSISI-source}. As this system of equations can be written in the form $\bm{L}e\bm{\alpha}=\bm{R} \bm{\alpha}$, the convergence rate of original GSIS can be determined by calculating the spectral radius of the matrix $\bm{G}=\bm{L}^{-1}\bm{R}$. 
As shown in Fig.~\ref{fig:converge_CIS_GSIS}, original GSIS exhibits excellent performance in the low-frequency, near-continuum regime, i.e., as the Knudsen number decreases, the spectral radius rapidly approaches zero, implying very fast convergence in near-continuum flows. 
However, for $\mathrm{Kn}=10$, two narrow ill-conditioned regions appear around $S\approx 0.7$ and $S\approx 1.3$, where the spectral radius even exceeds unity, indicating that the original GSIS may become unstable and fail to converge.

The first peak near $S\approx 0.7$ can be qualitatively understood from the structure of $\bm{L}$. 
In particular, one diagonal entry contains the term
$2\imath{S}+1/(\imath{S}+\delta_{rp})$.
When $\mathrm{Kn}$ is large,  it approaches $\imath(2{S}-1/{S})$. Thus, when $2S\approx 1/S$,  a marked reduction of the diagonal magnitude is seen. 
Consequently, $\bm{L}$ becomes closer to singular (its smallest singular value decreases and the condition number increases), which  produces the first sharp peak in the eigenvalues of $\bm{G}$.

The second peak around $S\approx 1.3$ cannot be attributed to a single diagonal entry; instead, it arises from the coupled multi-variable structure of the $6\times 6$ system. 
In this frequency range, the interaction among the density, momentum, energy, and heat flux can significantly reduce a certain singular value of $\bm{L}$, resulting in a rapid growth of $\mathrm{cond}(\bm{L})$ and, in turn, an elevated spectral radius of $\bm{G}$.

In Ref.~\cite{su2020can}, original GSIS was applied to one-dimensional oscillatory Couette flow and acoustic-wave problems. When both the Knudsen and the Strouhal numbers are large, the computations may be inaccurate or even diverge. To mitigate this issue, a relaxation term was added to the diagonal entries of the discrete system to enhance numerical stability. However, such a straightforward stabilization strategy may not be sufficiently robust for more complex configurations and can still lead to instability or inaccurate solutions. Therefore, to avoid these potential issues, the original GSIS must be modified further to guarantee stable and reliable convergence.

\subsection{Frequency-domain GSIS}
The ill-conditioning of the original GSIS can be traced to the asymmetric use of the NSF decomposition in Eq.~\eqref{eq:stress_heat_evolve0}.
Specifically, the stress and heat flux on the right-hand side of the equation are decomposed using linear constitutive relations, whereas those on the left-hand side are not. To fix this issue, we rearrange Eq.~\eqref{eq:stress_heat_evolve} into the following symmetric form:
\begin{equation*}
	\begin{aligned}
	&	\imath{S} \left(\bm{\Pi} -\bm{\Pi}_\text{NSF}+\bm{\Pi}^{n+\frac{1}{2}}_\text{NSF} \right) + 
		\int 2\left(\bm{v}\bm{v}-\frac{v^2}{3}\bm{I}\right)\,\bm{v}\cdot\nabla h^{n+\frac{1}{2}}f_{eq}d\bm{v}
		= -\delta_{rp}\left(\bm{\Pi} -\bm{\Pi}_\text{NSF}+\bm{\Pi}^{n+\frac{1}{2}}_\text{NSF} \right), \\
	&	\imath{S} \left( \bm{q}-\bm{q}_\text{NSF} +\bm{q}^{n+\frac{1}{2}}_\text{NSF} \right) +
		\int \left(v^2-\frac{5}{2}\right)\bm{v}(\bm{v}\cdot\nabla h^{n+\frac{1}{2}}) f_{eq}d\bm{v}
		= -\frac{2}{3}\delta_{rp}
        \left( \bm{q}-\bm{q}_\text{NSF} +\bm{q}^{n+\frac{1}{2}}_\text{NSF} \right).
	\end{aligned}
\end{equation*}
Then by decomposing the stress and heat flux into the linear NSF constitutive relations and the high-order terms as 
\begin{equation}\label{modified_GSIS}
 \begin{aligned}
    \bm{\Pi}=& \bm{\Pi}_\text{NSF} + \mathbf{HoT}^{n+\frac{1}{2}}_{\bm{\Pi}}, \\
    \bm{q} =& \bm{q}_\text{NSF} +  \mathbf{HoT}^{n+\frac{1}{2}}_{\bm{q}},
 \end{aligned}
\end{equation}
we have 
\begin{equation}\label{eq:modified GSIS-I-HoT}
    \begin{aligned}
&   \mathbf{HoT}^{n+\frac{1}{2}}_{\bm{\Pi}}   
    = -\frac{1}{\delta_{rp}+\imath{S}}
    \int 2\left(\bm{v}\bm{v}-\frac{v^2}{3}\bm{I}\right)
    (\bm{v} \cdot \nabla h^{n+\frac{1}{2}}) 
    f_{eq}d\bm{v} - \bm{\Pi}^{n+\frac{1}{2}}_\text{NSF}, \\
  &  \mathbf{HoT}^{n+\frac{1}{2}}_{\bm{q}} 
    = -\frac{1}{\frac{2}{3}\delta_{rp}+\imath{S}}
    \int \left(v^2-\frac{5}{2}\right)\bm{v}
    (\bm{v} \cdot \nabla h^{n+\frac{1}{2}})
    f_{eq}d\bm{v} - \bm{q}^{n+\frac{1}{2}}_\text{NSF}.
    \end{aligned}
\end{equation}

The final linear system of equations for $\bm{\alpha}$ is given by
\begin{equation}\label{eq:GSIS-F-Matrix}
    \begin{aligned}
    &    e(\imath{S}\alpha_\rho
    +\imath\bm{\theta} \cdot \alpha_{\bm{u}})=0, \\
    &e\left[\imath\theta_x(\alpha_\rho+\alpha_\tau)+
    \left(2\imath{S}
    +\delta_{rp}^{-1}\right)\alpha_{u_x}\right]=S_2, \\
    &e\left[\imath\theta_y(\alpha_\rho+\alpha_\tau)+
    \left(2\imath{S}
    +\delta_{rp}^{-1}\right)\alpha_{u_y} \right]=S_3, \\
     &   e\left(\frac{3}{2}\imath{S}\alpha_\tau
     + \imath \bm{\theta} \cdot \alpha_{\bm{u}} 
     + \imath \bm{\theta} \cdot \alpha_{\bm{q}}\right)=0, \\
     &   e\left(\frac{15}{8}\imath\theta_x\delta^{-1}_{rp} \alpha_\tau 
        +\alpha_{q_x}\right)=S_5,   \\
      &  e\left(\frac{15}{8}\imath\theta_y \delta^{-1}_{rp} \alpha_\tau
        +\alpha_{q_y}\right)=S_6,   \\
    \end{aligned}
\end{equation}
where the source terms, due to the high-order terms, are expressed as
\begin{equation}\label{eq:modified GSIS-I-source}
    \begin{aligned}
   & S_2=\int \left[
    \delta^{-1}_{rp}v_x-\frac{2\bm{\theta} \cdot \bm{v}}{\imath{S}+\delta_{rp}}\left(
    \theta_x\left(v_x^2-\frac{v^2}{3}\right)+\theta_y v_x v_y
    \right)\right]\bar{Y}(\bm{v})d\bm{v}, \\
  &  S_3=\int \left[
    \delta^{-1}_{rp}v_y -\frac{2\bm{\theta} \cdot \bm{v}}{\imath{S}+\delta_{rp}}\left(
    \theta_y\left(v_y^2-\frac{v^2}{3}\right)+\theta_x v_x v_y
    \right)\right]\bar{Y}(\bm{v})d\bm{v}, \\
    & S_5=\int\left[
        \frac{15}{8}\imath\delta^{-1}_{rp}\theta_x\left(\frac{2}{3}v^2-1\right)
        -\frac{\imath\bm{\theta} \cdot \bm{v}}{\imath{S}+\frac{2}{3}\delta_{rp}}v_x\left(v^2-\frac{5}{2}\right)
    \right]\bar{Y}(\bm{v})d\bm{v},  \\
   & S_6=\int\left[
        \frac{15}{8}\imath\delta^{-1}_{rp}\theta_y\left(\frac{2}{3}v^2-1\right)
        -\frac{\imath\bm{\theta} \cdot \bm{v}}{\imath{S}+\frac{2}{3}\delta_{rp}}v_y\left(v^2-\frac{5}{2}\right)
    \right]\bar{Y}(\bm{v})d\bm{v}.
    \end{aligned}
\end{equation}

The spectral radius of the frequency-domain GSIS is shown in Fig.~\ref{fig:converge_CIS_GSIS}. Compared with the original GSIS, the "spike" is eliminated. In the near-continuum regime, the frequency-domain GSIS maintains a small spectral radius over the entire frequency range, indicating robust and fast convergence. More importantly, since the spectral radius scales as $1/\delta_{rp}^2$, the false convergence of the CIS in the continuum flow regime is transformed into super-convergence for the GSIS~\cite{su2020fast}. This follows from
\begin{equation} \label{GSIS_super}
   \frac{\rho(\bm{G})}{1-\rho(\bm{G})} \epsilon 
   \rightarrow  \frac{\epsilon}{\delta_{rp}^2}, ~\text{when}~ \delta_{rp}\rightarrow\infty.
\end{equation}
Consequently, even when the convergence criterion $\epsilon$ is relatively large, the GSIS can still rapidly obtain an accurate solution.

Therefore, in the following paper, the frequency-domain GSIS is adopted in all subsequent computations. In the remainder, GSIS refers to this frequency-domain variant unless stated otherwise.

\subsection{Asymptotic-preserving property of GSIS}

The AP property implies that, in the presence of weak rarefaction effects, the numerical solution of the kinetic equation coincides with that of the NSF equations. Su \textit{et al.}~\cite{su2020fast} have demonstrated the AP property of GSIS for steady-state problems. Here we analyze the frequency domain GSIS. Prior to this analysis, it is noteworthy that rarefaction effects in time-independent problems are solely characterized by the spatial Knudsen number ($\delta^{-1}_{rp}$), whereas in time-dependent scenarios, they are additionally governed by the Strouhal number 
$S$, or more precisely, the temporal Knudsen number $Kn_t$ \cite{LeiBook}. We herein assume  $Kn, Kn_t\ll1$, such that the continuum assumption of gas dynamics holds.

When the iteration converges, the discrete form of the kinetic equation~\eqref{eq:dvmeqn} can be rewritten as
\begin{equation}\label{discrete_S}
    \imath{S} h + \bm{v} \cdot \nabla_\delta h = Lh, 
    \quad \text{with} \quad 
    \nabla_\delta = \nabla+ O(\Delta x^{n}),
\end{equation}
where $n$ is the order of spatial accuracy
and $\Delta x$ is the mesh size.
Substituting the Taylor expansion
$h = h_0 + \delta_{rp}^{-1} h_1 + \delta_{rp}^{-2} h_2 + ...$  into Eq.~\eqref{discrete_S}, and collecting terms of the order $\delta_{rp}^{-1},\delta_{rp}^{0}$, respectively, we obtain the following equations when $\Delta x\sim O(1)$: 
\begin{equation}\label{CE_modified}
\begin{aligned}
    h_0=&\rho + 2\bm{v} \cdot \bm{u}
    + \left(v^2 - \frac{3}{2}\right)\tau
    + \frac{4}{15}\left(v^2 - \frac{5}{2}\right)\bm{v} \cdot \bm{q},\\
   h_1=&- \bm{v} \cdot \nabla_\delta h_0. 
\end{aligned}
\end{equation}
It should be noted that, in the normal Chapman-Enskog expansion, the $h_1$ term is used to produce the linear NSF constitutive relations. However, in the frequency-domain GSIS, the constitutive relation is constructed as Eq.~\eqref{eq:modified GSIS-I-HoT}.

With Eq.~\eqref{CE_modified},  the high-order term in the heat flux is calculated as
\begin{equation}
\begin{aligned}[b]
\mathbf{HoT}_{\bm{q}}
    =&-\frac{1}{\frac{2}{3}\delta_{rp}+\imath{S}}
    \int \left(v^2-\frac{5}{2}\right)\bm{v}
    (\bm{v} \cdot \nabla h)
    f_{eq}d\bm{v} - \bm{q}_{\text{NSF}} \\
    =& -\frac{1}{\frac{2}{3}\delta_{rp}+\imath{S}}
    \int \left(v^2-\frac{5}{2}\right)\bm{v}
    \left[\bm{v} \cdot \nabla (h_0+\delta_{rp}^{-1}h_1)\right]
    f_{eq}d\bm{v} - \bm{q}_{\text{NSF}} +O(\delta^{-2}_{rp})
    \\
    =&  -\frac{3}{2\delta_{rp}}
    \int \left(v^2-\frac{5}{2}\right)\bm{v}
    \left[\bm{v} \cdot \nabla h_0
    \right]
    f_{eq}d\bm{v} - \bm{q}_{\text{NSF}}  
    \\
    &+\frac{3}{2\delta^2_{rp}}
    \int \left(v^2-\frac{5}{2}\right)\bm{v}
    \left[ \bm{v} \cdot \nabla (\bm{v}\cdot \nabla_\delta h_0)\right]
    f_{eq}d\bm{v}  +O(\delta^{-2}_{rp})
    \\
    =&O(\delta^{-2}_{rp}).
\end{aligned}
\end{equation}

The same applied to the high-order terms in stress.
Therefore, according to Eq.~\eqref{modified_GSIS}, the NSF equations are recovered when $\Delta x \sim O(1)$ in the continuum regime, as long as this size is about to resolve the hydrodynamic flow. 
In contrast, if the heat flux is calculated directly according to its definition \eqref{eq:moment}, then Eq.~\eqref{CE_modified} yields $\bm{q}=-(15/8\delta_{rp})\nabla_\delta \tau$, which represents the Fourier heat conduction law affected by spatial discretization errors. Consequently, the scheme is asymptotic-preserving to the NSF limit only when $\Delta x\sim O(Kn^{1/n})$.


\section{Numerical methods}
\label{sec:numerical}

This section presents the numerical method for solving the frequency-domain linearized kinetic equation \eqref{eq:dvmeqn}. We first introduce the finite-volume discretization for both the kinetic equation and the macroscopic synthetic equations. We then detail the resulting linear algebraic formulation and the solution strategy, and finally provide an overview of the GSIS framework.

\subsection{Conventional iteration scheme}
A cell-centered finite-volume method (FVM) is employed to discretize the frequency-domain linearized Boltzmann equation. We use subscripts $i$ and $j$ to denote control volumes (cells), $ij$ to denote the face shared by cells $i$ and $j$. Let $V_i$ be the volume of cell $i$, and $\bm{S}_{ij}=S_{ij}\bm{n}_{ij}$ be the outward face-area vector pointing from cell $i$ to cell $j$, where $\bm{n}_{ij}$ is the unit normal and $S_{ij}=\|\bm{S}_{ij}\|$.

In the discrete velocity method, the molecular velocity $\bm{v}$ is discretized. For a fixed discrete velocity \(\bm{v_k}\), integrating Eq.~(\ref{eq:dvmeqn}) over the cell $i$ and applying Gauss' theorem yields
\begin{equation}
\label{eq:fv_kinetic}
    (\delta_{rp}+\imath{S})\,h_{i,k}^{\,n+1}
    +\frac{1}{V_i}\sum_{j\in N(i)} \mathcal{F}_{ij,k}^{\,n+1}
    = L_{i,k}^{\,n},
\end{equation}
where $N(i)$ denotes the neighbor set of cell $i$, the superscript $n$ denotes the outer iteration index, and the source term $L_{i,k}^{\,n}$ is evaluated from the macroscopic moments at iteration $n$:
\begin{equation}
\label{eq:L_shakhov}
    L_{i,k}^{\,n}=
    \delta_{rp}\!\left[
    \rho_i^{\,n}+2\bm{v}_k\!\cdot\!\bm{u}_i^{\,n}
    +\Big(v_k^2-\frac{3}{2}\Big)\tau_i^{\,n}
    +\frac{4}{15}\Big(v_k^2-\frac{5}{2}\Big)\bm{v}_k\!\cdot\!\bm{q}_i^{\,n}
    \right].
\end{equation}
The numerical flux $\mathcal{F}_{ij,k}$ in Eq.~\eqref{eq:fv_kinetic} is defined as
\begin{equation}
\label{eq:flux_def}
    \mathcal{F}_{ij,k}^{\,n+1}
    = \left(\xi_{ij,k}^{+}\,h_{ij,k}^{L,n+1}+\xi_{ij,k}^{-}\,h_{ij,k}^{R,n+1}\right)S_{ij},
    \quad
    \xi_{ij,k}=\bm{v}_k\cdot\bm{n}_{ij},
\end{equation}
with the standard splitting
\begin{equation}
\label{eq:xi_split}
    \xi_{ij,k}^{+}=\max(\xi_{ij,k},0),\quad
    \xi_{ij,k}^{-}=\min(\xi_{ij,k},0).
\end{equation}
Here $h_{ij,k}^{L}$ and $h_{ij,k}^{R}$ are the reconstructed values at face $ij$ from cells $i$ and $j$, respectively. A second-order linear reconstruction is used,
\begin{equation}\label{eq:reconstruction}
\begin{aligned}
  &  h_{ij,k}^{L,n+1}=h_{i,k}^{\,n+1}+\nabla h_{i,k}^{\,n}\cdot(\bm{x}_{ij}-\bm{x}_i),   
    \\
  &  h_{ij,k}^{R,n+1}=h_{j,k}^{\,n+1}+\nabla h_{j,k}^{\,n}\cdot(\bm{x}_{ij}-\bm{x}_j),
\end{aligned}
\end{equation}
where $\bm{x}_i$, $\bm{x}_j$ and $\bm{x}_{ij}$ denote the centroids of cells $i$, $j$ and face $ij$. The gradient $\nabla h_{i,k}$ is computed by a least-squares procedure based on neighboring cell values.

Equation~\eqref{eq:fv_kinetic} can be written in the standard sparse-matrix form
\begin{equation}
\label{eq:matrix_kinetic}
    d_i\,h_{i,k}^{\,n+1}+\sum_{j\in N(i)} d_{ij}\,h_{j,k}^{\,n+1}=b_{i,k}^{\,n},
\end{equation}
where $d_i$ and $d_{ij}$ denote the diagonal and off-diagonal coefficients induced by the upwind flux \eqref{eq:flux_def}, and $b_{i,k}^{\,n}$ collects all explicit terms (including reconstruction-gradient contributions). The linear system is solved by a lower--upper symmetric Gauss--Seidel iteration \cite{yoon1988lower}.





\subsection{Macroscopic synthetic equations}
\label{subsec:simple_coupled}

The macroscopic synthetic equations can be expressed as
\begin{equation}
\label{eq:synthetic_cont_mom_energy}
\begin{aligned}
 &   \imath{S}\,\rho + \nabla\cdot\bm{u} = 0, \\[2pt]
 &   2\imath{S}\,\bm{u} + \nabla(\rho+\tau) + \nabla\cdot\bm{\Pi}_\text{NSF}
    = -\nabla\cdot \mathbf{HoT}_{\bm{\Pi}}^{n+\frac{1}{2}}, \\[2pt]
 &   \frac{3}{2}\imath{S}\,\tau + \nabla\cdot\bm{u} + \nabla\cdot\bm{q}_\text{NSF}
    = -\nabla\cdot \mathbf{HoT}_{\bm{q}}^{n+\frac{1}{2}},
\end{aligned}
\end{equation}

When $S=0$ and $\delta_{rp}\gg 1$, Eq.~\eqref{eq:synthetic_cont_mom_energy} degenerates to a Stokes-type saddle-point system in which the pressure-like variable $p=\rho+\tau$ acts as a Lagrange multiplier enforcing mass conservation. 
For $S>0$, the terms proportional to $S$ provide an additional temporal coupling among $\rho$, $\boldsymbol{u}$, and $\tau$, so that the macroscopic variables must be advanced in a tightly coupled manner. 
To ensure robustness across the whole range of $(S,\delta_{rp})$ and to avoid stringent stability restrictions, it is solved by a fully implicit block-coupled formulation~\cite{darwish2009coupled}.

\subsubsection{Collocated arrangement and Rhie--Chow interpolation}
All macroscopic unknowns are stored at cell centers. To avoid checkerboard decoupling in the continuity equation, the face-normal flux is computed using a Rhie--Chow-type interpolation \cite{rhie1983numerical}. Define
\begin{equation}
\label{eq:p_def}
    p \equiv \rho+\tau.
\end{equation}
Let $\overline{\phi_{ij}}$ denote a linear interpolation to face $ij$,
\begin{equation}
\label{eq:linear_interp}
    \overline{\phi_{ij}} = g_{ij}\phi_i + (1-g_{ij})\phi_j,
\end{equation}
where $g_{ij}\in[0,1]$ is determined from the face location along the segment connecting $\bm{x}_i$ and $\bm{x}_j$.
The Rhie--Chow corrected face velocity is written as
\begin{equation}
\label{eq:rhie_chow}
    \bm{u}_{ij}\cdot\bm{S}_{ij}
    =
    \overline{\bm{u}}_{ij}\cdot\bm{S}_{ij}
    -
    \overline{\bm{D}}_{ij}
    \left(\nabla p_{ij}-\overline{\nabla p}_{ij}\right)\cdot\bm{S}_{ij},
    \quad
    \bm{D}_i=\frac{V_i}{a_i^{\bm{u}}},
\end{equation}
where $a_i^{\bm{u}}$ is the main diagonal coefficient of the discretized momentum equation (for each velocity component), and $\bm{D}_i$ plays the role of an inverse momentum operator.

The face gradient of $p$ is approximated by a consistent decomposition:
\begin{equation}
\label{eq:gradp_face}
\nabla p_{ij}
=
\overline{\nabla p}_{ij}
+
\left[
\frac{p_j-p_i}{d_{ij}}
-
\left(\overline{\nabla p}_{ij}\cdot\bm{e}_{ij}\right)
\right]\bm{e}_{ij},
\quad
d_{ij}=\|\bm{x}_j-\bm{x}_i\|,
\quad
\bm{e}_{ij}=\frac{\bm{x}_j-\bm{x}_i}{d_{ij}}.
\end{equation}

\subsubsection{Finite-volume discretization of the synthetic equations}

Let $m$ denote the inner-iteration index used to explicitly treated terms; we initialize $(\rho,\bm{u},\tau)^{m=0}$ from the kinetic half-step moments. 
Integrating Eq.~\eqref{eq:synthetic_cont_mom_energy} over cell $i$ and applying Gauss' theorem yields
\begin{equation}
\label{eq:fv_synthetic}
\begin{aligned}
    &\imath{S}\,V_i \rho_i^{m+1}
    +
    \sum_{j\in N(i)}
    \left(\bm{u}_{ij}^{m+1}\cdot\bm{S}_{ij}\right)
    =0,\\[2pt]
    &2\imath{S}\,V_i \bm{u}_i^{m+1}
    +\sum_{j\in N(i)} \overline{p}_{ij}^{\,m+1}\,\bm{S}_{ij}
    -\frac{1}{\delta_{rp}}\sum_{j\in N(i)}
    \left(\nabla \bm{u}_{ij}^{m+1}\cdot\bm{S}_{ij}\right)
    =\bm{b}_i^{\bm{u}},\\[2pt]
    &\frac{3}{2}\imath{S}\,V_i \tau_i^{m+1}
    +\sum_{j\in N(i)}\left(\bm{u}_{ij}^{m+1}\cdot\bm{S}_{ij}\right)
    -\frac{15}{8\delta_{rp}}\sum_{j\in N(i)}
    \left(\nabla \tau_{ij}^{m+1}\cdot\bm{S}_{ij}\right)
    =b_i^{\tau},
\end{aligned}
\end{equation}
where the right-hand sides collect the explicitly treated terms:
\begin{equation}
\label{eq:rhs_synthetic}
\begin{aligned}
  &  \bm{b}_i^{\bm{u}}
    =
    -\sum_{j\in N(i)}
    \left[
    \frac{1}{\delta_{rp}}
    \left(
    \big((\nabla\bm{u})_{ij}^{m}\big)^{T}
    -\frac{2}{3}(\nabla\cdot\bm{u})_{ij}^{m}\bm{I}
    \right)
    +\left(\mathbf{HoT}_{\bm{\Pi}}^{n+\frac{1}{2}}\right)_{ij}
    \right]\cdot\bm{S}_{ij},\\[3pt]
  &  b_i^{\tau}
    =
    -\sum_{j\in N(i)}
    \left(\mathbf{HoT}_{\bm{q}}^{n+\frac{1}{2}}\right)_{ij}\cdot\bm{S}_{ij}.
\end{aligned}
\end{equation}
where the high order terms in the face is evaluated as per Eq.~\eqref{eq:linear_interp}.

For the diffusive term on a non-orthogonal mesh, we adopt the over-relaxed correction.
Let $\bm{d}_{ij}=\bm{x}_j-\bm{x}_i$, $d_{ij}=\|\bm{d}_{ij}\|$, and $\bm{e}_{ij}=\bm{d}_{ij}/d_{ij}$.
Define the over-relaxed orthogonal contribution
\begin{equation}
\label{eq:overrelaxed_Delta_T}
\bm{E}_{ij}=\frac{\|\bm{S}_{ij}\|^{2}}{\bm{S}_{ij} \cdot \bm{e}_{ij}}\,\bm{e}_{ij},
\quad
\bm{T}_{ij}=\bm{S}_{ij}-\bm{E}_{ij}.
\end{equation}
Then the diffusive flux is approximated by an implicit main term plus an explicit non-orthogonal correction:
\begin{equation}
\label{eq:diff_overrelaxed_compact}
\nabla \phi_{ij}^{m+1}\cdot\bm{S}_{ij}
\approx
\underbrace{
\frac{\|\bm{E}_{ij}\|}{d_{ij}}
\left(\phi_{j}^{m+1}-\phi_{i}^{m+1}\right)}_{\text{implicit main term}}
+
\underbrace{\overline{\nabla \phi}_{ij}^{\,m}\cdot\bm{T}_{ij}}_{\text{explicit correction}},
\end{equation}
where $\phi\in\{\rho,u_x,u_y,\tau\}$ and $\overline{\nabla \phi}_{ij}^{\,m}$ is obtained by linear interpolation of cell-centered gradients.

\subsubsection{Coupled linear system}
In the coupled algorithm, the density, velocity and temperature equations are assembled into a single block system for the unknown vector
\(
\bm{W}=[\rho,u_x,u_y,\tau]^\top\).
The resulting algebraic system is sparse and complex-valued due to the $\imath{S}$ terms. For a fixed mesh and fixed parameters $(\delta_{rp},S)$, the coefficient matrix does not change across iterations. Therefore, we assemble the matrix once and compute its sparse LU factorization using \texttt{Eigen::SparseLU}~\cite{eigenweb}. At each iteration, only the right-hand side is updated, and the synthetic solution is obtained by forward/back substitution. This strategy avoids expensive matrix inner iterations and significantly reduces the macroscopic-solve cost.

\subsection{Overview of GSIS}
In summary, the iterative procedure of GSIS can be summarized as follows.
\begin{enumerate}
    \item At the $n$-th iteration, the velocity distribution function $h^n$ and the macroscopic quantities $\bm{W}^n=[\rho^n,\bm{u}^n,\tau^n,\bm{\Pi}^n,\bm{q}^n]$ are known. Solve Eq.~(\ref{eq:dvmeqn}) to obtain the intermediate distribution function $h^{n+\frac{1}{2}}$.

    \item Compute macroscopic moments from $h^{n+\frac{1}{2}}$ using Eq.~(\ref{eq:moment}). The high-order terms are then evaluated using
    Eq.~(\ref{eq:modified GSIS-I-HoT}).

    \item Substitute the high-order terms into the macroscopic synthetic system~(\ref{eq:sysEqn}) and solve for the updated macroscopic variables $\bm{W}^{n+1}$. In the present implementation, this step is performed using the block-coupled formulation described in Section~\ref{subsec:simple_coupled}.

    \item Correct the distribution function using the updated macroscopic quantities:
    \begin{equation}
        \label{update_vdf}
        h^{n+1}=h^{n+\frac{1}{2}}
        +\lambda_{\rho}
        +2\bm{v}\cdot\lambda_{\bm{u}}
        +\left(v^2-\frac{3}{2}\right)\lambda_{\tau},
    \end{equation}
    where $\lambda_{\rho}=\rho^{n+1}-\rho^{n+1/2}$, 
    $\lambda_{\bm{u}}=\bm{u}^{n+1}-\bm{u}^{n+1/2}$, 
    and $\lambda_{\tau}=\tau^{n+1}-\tau^{n+1/2}$.
\end{enumerate}

The above steps are repeated until the residuals of macroscopic variables fall below a prescribed tolerance.

\section{Result and discussion}
\label{sec:result}

In this section, we present two representative numerical examples to assess the accuracy and efficiency of the proposed GSIS method for linear oscillatory gas flows. For both CIS and GSIS, convergence is declared when the following relative residual falls below a prescribed tolerance:
\begin{equation}\label{eq:residual}
    \epsilon=\left[\frac{\int_{\Omega}\left(\phi^{\,n+1}-\phi^{\,n}\right)^2\,dx\,dy}{\int_{\Omega}\left(\phi^{\,n}\right)^2\,dx\,dy}\right]^{1/2},
    \quad 
    \phi=\big[\rho,\,u_x,\,u_y,\,\tau\big],
\end{equation}
where \(\Omega\) denotes the computational domain and \(n\) is the iteration index. Unless otherwise specified, the convergence tolerance is set to \(10^{-5}\).



\subsection{Oscillatory shear-driven flow between two eccentric cylinders}
\label{sec:cylinders}

\begin{figure}
    \centering
    \includegraphics[width=0.4\linewidth]{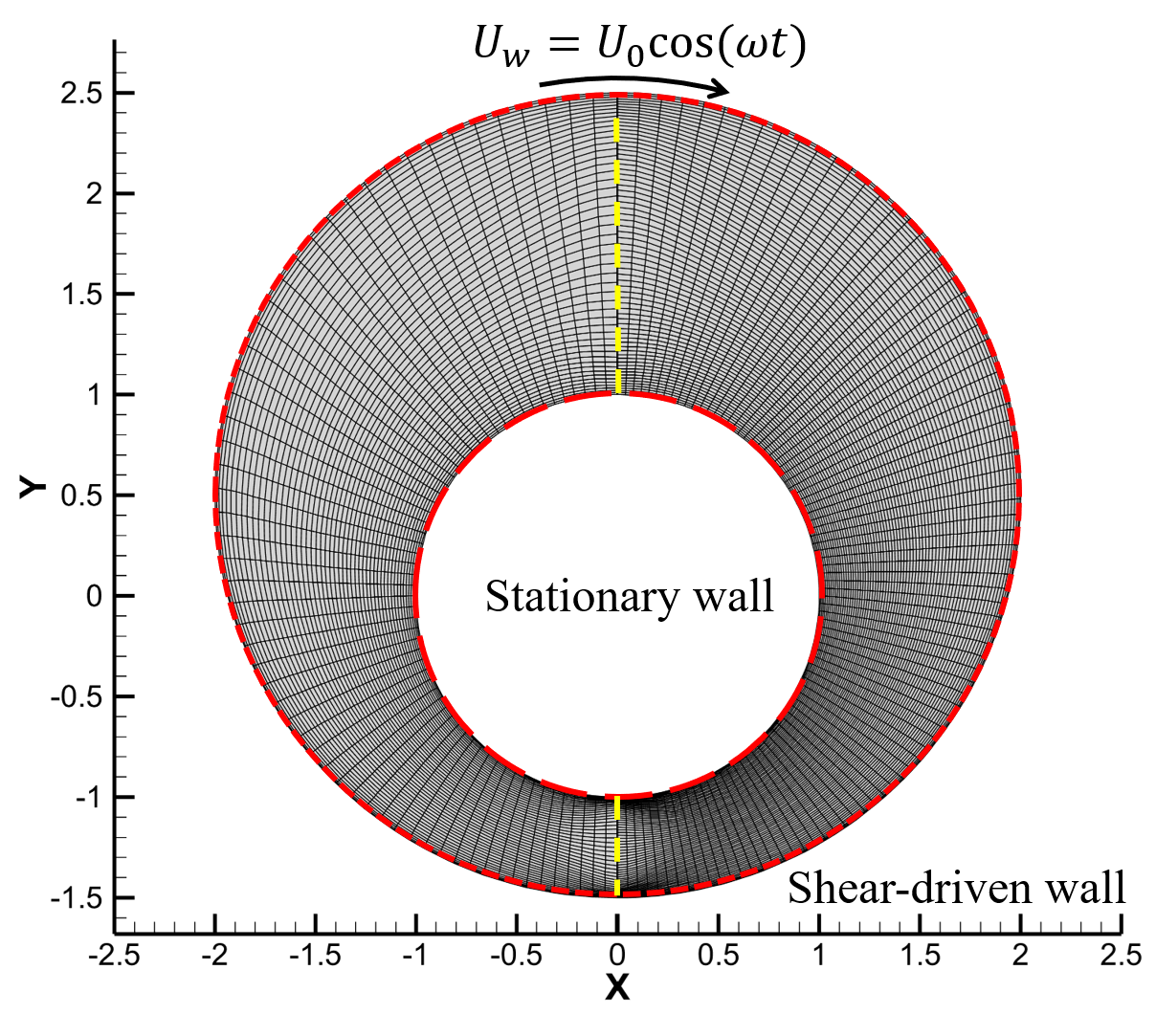}\\ \vspace{0.5cm}
    {\includegraphics[width=0.4\linewidth]
    {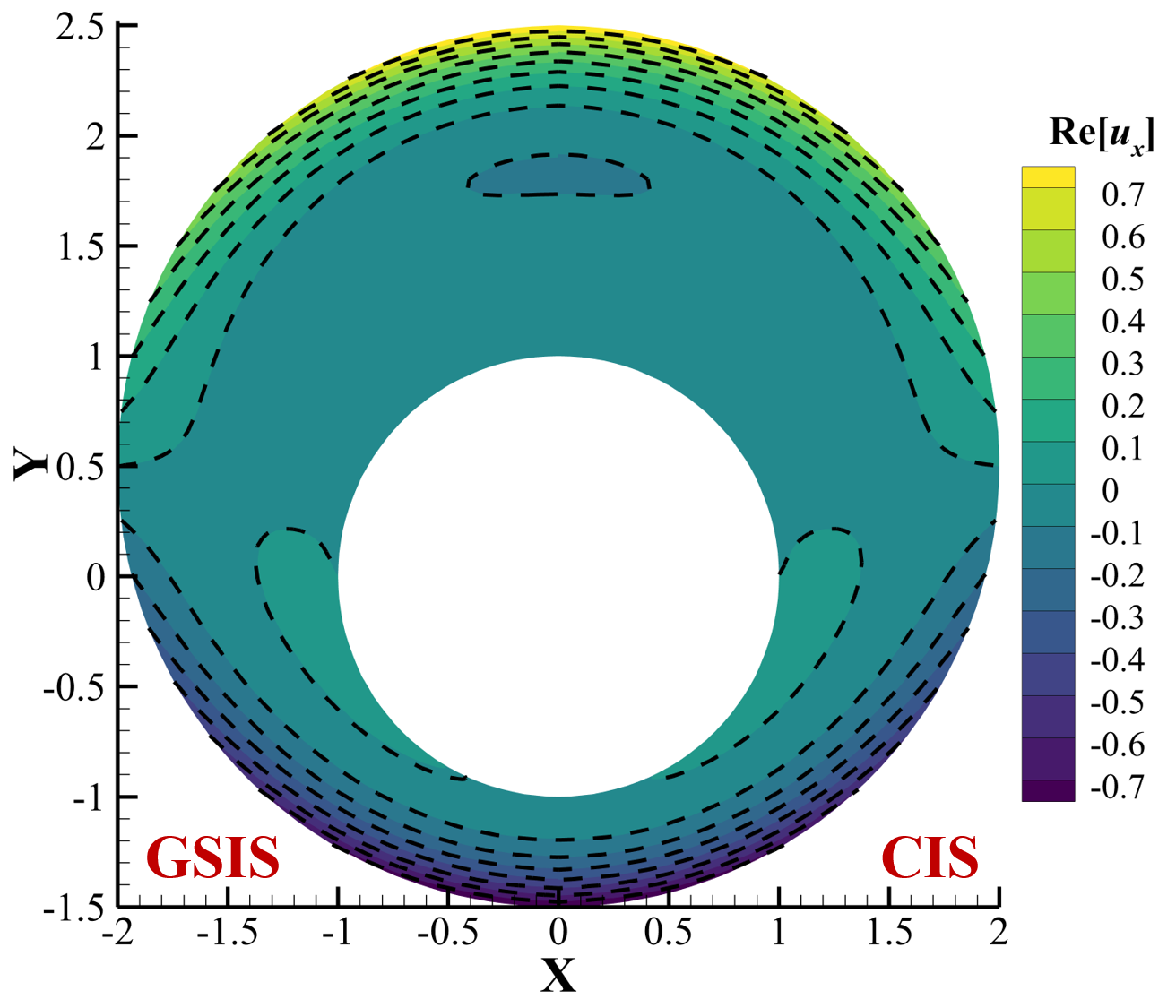}}
    \hspace{0.5cm}
    {\includegraphics[width=0.4\linewidth]
    {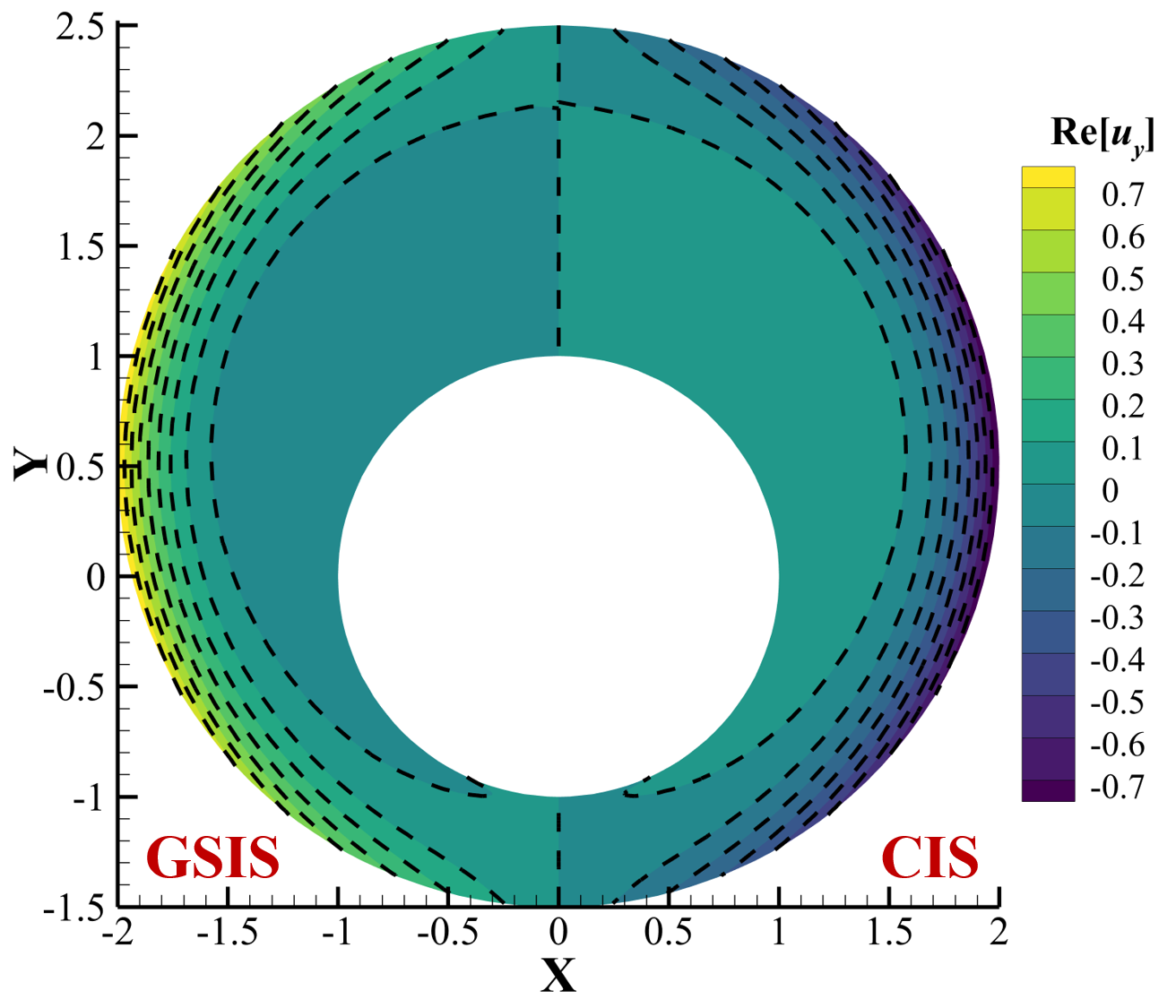}}\\ \vspace{0.5cm}
    {\includegraphics[width=0.4\linewidth]
    {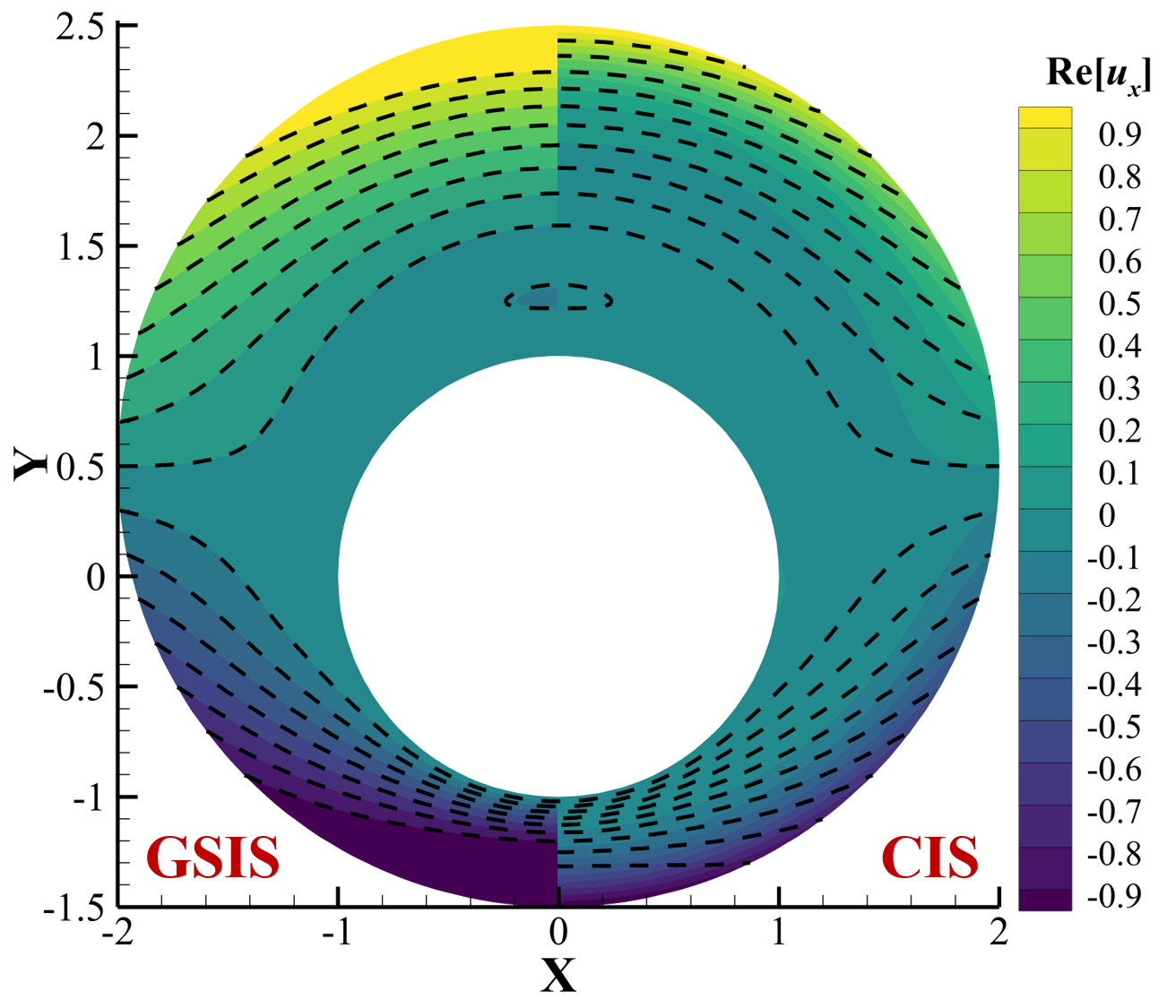}}
    \hspace{0.5cm}
    {\includegraphics[width=0.4\linewidth]
    {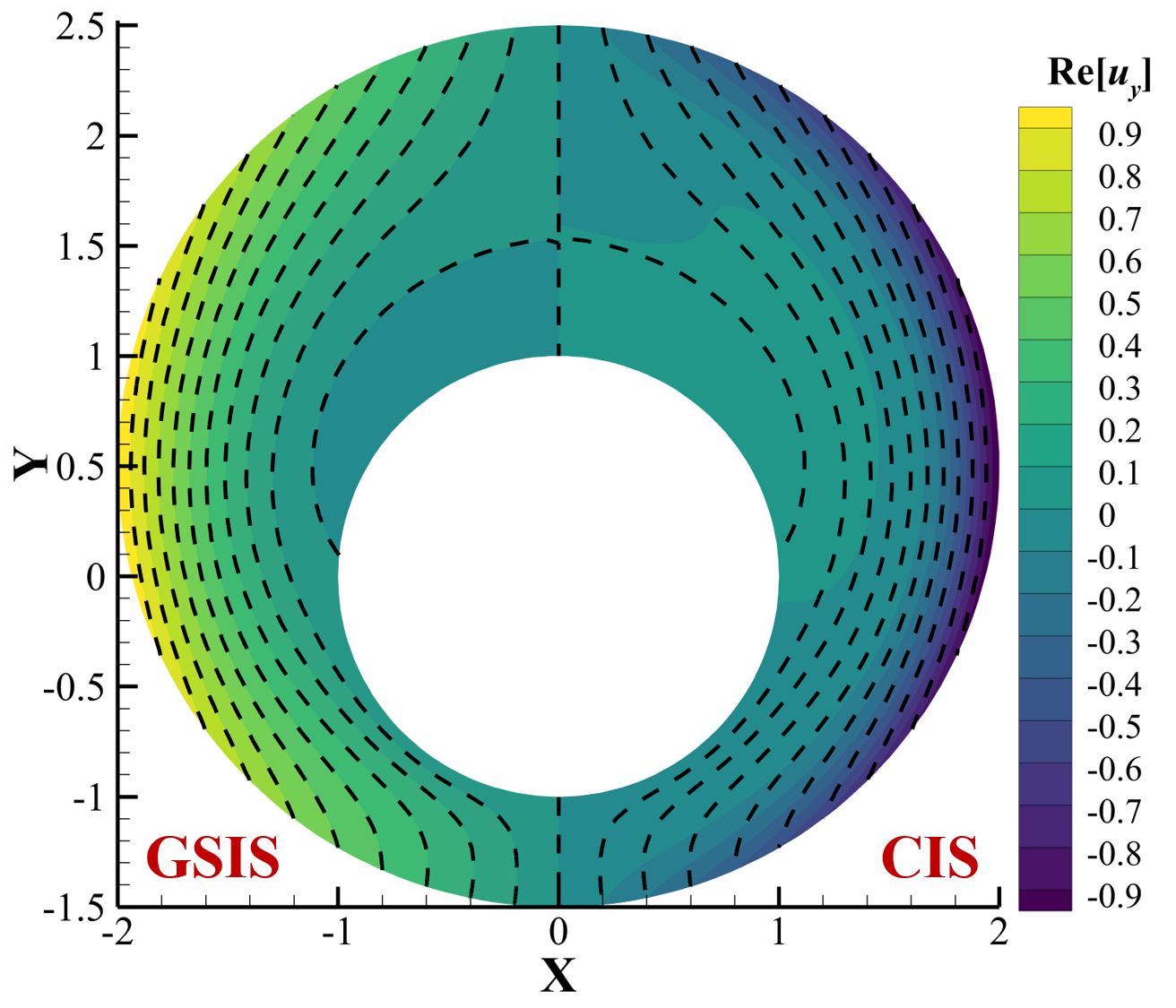}}
    \caption{
    Oscillatory shear-driven flow between two eccentric cylinders.
    \textbf{Top:} Geometry and structured mesh. 
    The yellow dashed lines indicate the locations where velocity profiles are extracted. 
    Two meshes with $N_{\text{cell}}\approx 6,000$ (Left) and $12,000$ (Right) are employed.
    \textbf{Middle:} Contours of $\Re(u_x)$ and $\Re(u_y)$ for $\delta_{rp}=10$ and $S=1$.
    \textbf{Bottom:} Contours of $\Re(u_x)$ and $\Re(u_y)$ for $\delta_{rp}=1000$ and $S=0.001$. 
    The black dashed lines denote the reference solution used for comparison.
    }
    \label{fig:cylinder}
\end{figure}

Consider an oscillatory flow between two eccentric cylinders~\cite{su2020can}.
The outer cylinder of radius 2 oscillates with tangential velocity $U_w(t)=U_0\cos(\omega t)$,
while the inner cylinder of radius 1 remains stationary.
The radius of the inner cylinder is used as the reference length.
The centers of the outer and inner cylinders are located at $(0,0.5)$ and $(0,0)$, respectively, leading to a non-uniform annular gap. Figure~\ref{fig:cylinder} shows the structured meshes. To resolve the Knudsen layers, the mesh is refined toward solid surfaces with minimum and maximum grid spacings of $0.001\,L$ and \(0.1\,L\), respectively.
Two meshes are employed: $N_{\text{cell}}\approx 6,000$ for cases with $\delta_{rp}\le 10$ and $N_{\text{cell}}\approx 12,000$ for $\delta_{rp}\ge 100$.



The reduced-distribution formulations are employed
\begin{equation}
h_1(x,y,v_x,v_y)=\int_{-\infty}^{\infty} h(x,y,\bm{v})\,dv_z,\quad
h_2(x,y,v_x,v_y)=\int_{-\infty}^{\infty} v_z^2 h(x,y,\bm{v})\,dv_z,
\end{equation}
so that only $(v_x,v_y)$ are discretized.
A non-uniform grid is adopted for each component $v_\alpha$ ($\alpha\in\{x,y\}$):
\begin{equation}\label{eq:velocity_grid}
    v_{\alpha}^{(k)} = V_{\max}\left(\frac{2k-(N_{v,\alpha}+1)}{N_{v,\alpha}-1}\right)^{3},\quad
    k=1,2,\cdots,N_{v,\alpha},
\end{equation}
which clusters discrete velocities near $v_\alpha=0$ to capture the large variation of distribution function when the Knudsen number is large~\cite{wu2014oscillatory}.
In this work we use $N_{v,x}=N_{v,y}=32$ and $V_{\max}=6$.

For diffuse reflection at the moving outer cylinder, the reflected perturbation distribution satisfies
\begin{equation}
h = \frac{2}{\pi}\int_{v_n>0} v_n e^{-v^2}\,h\,d\bm{v} + 2\,\bm{t}_w\cdot \bm{v}, \quad v_n<0,
\end{equation}
where $\bm{t}_w$ is the outward unit tangential vector on the wall and $v_n= \bm{v} \cdot \bm{n}$ is the molecular velocity component along the outward wall normal.
The boundary condition at the inner cylinder is defined similarly but without the tangential driving term $2\,\bm{t}_w\cdot\bm{v}$.

\subsubsection{Asymptotic-preserving and super-convergence}

For the rarefied condition \((\delta_{rp},S)=(10,1)\), we compute both the CIS and GSIS solutions on the \(6,000\) mesh and take the CIS solution on the \(12,000\) mesh as the reference. As shown in Fig.~\ref{fig:cylinder}, GSIS excellently reproduces the CIS flow field: the flow velocities are nearly indistinguishable, indicating an essentially identical global flow structure. This close match confirms that GSIS is accurate in the rarefied regime.

We then consider the case \((\delta_{rp},S)=(1000,0.001)\) and compute both CIS and GSIS on the \(12,000\) mesh. Since the flow is continuum and the oscillation is quasi-steady, rarefaction effects are negligible; we therefore use the NSF solution on the same mesh as the reference. As shown in Fig.~\ref{fig:cylinder}, GSIS matches the reference well, demonstrating that it possesses the AP property.

\begin{figure}[t!]
    \centering
    \subfigure
    {\includegraphics[width=0.48\linewidth]
    {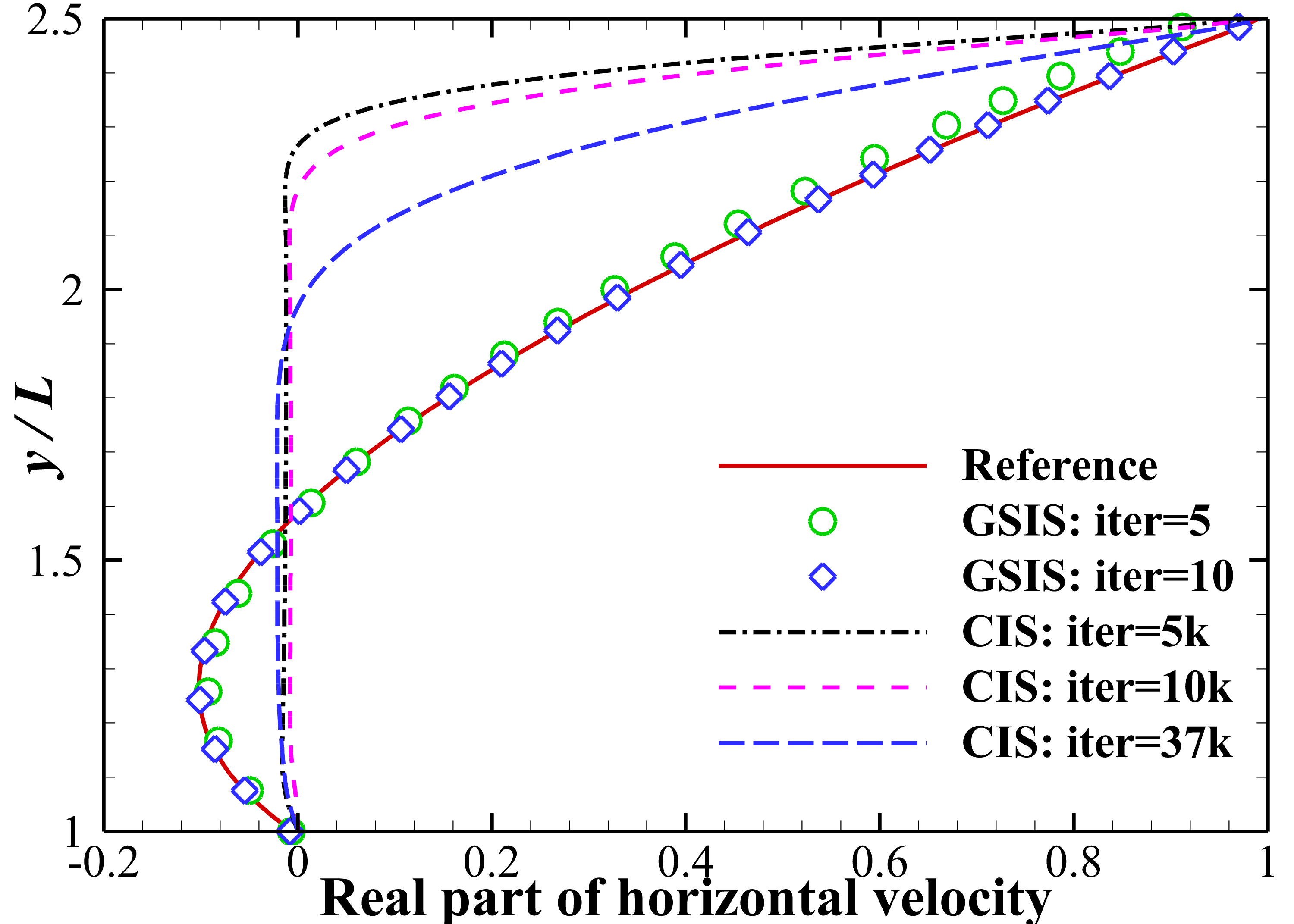}}
    {\includegraphics[width=0.48\linewidth]
    {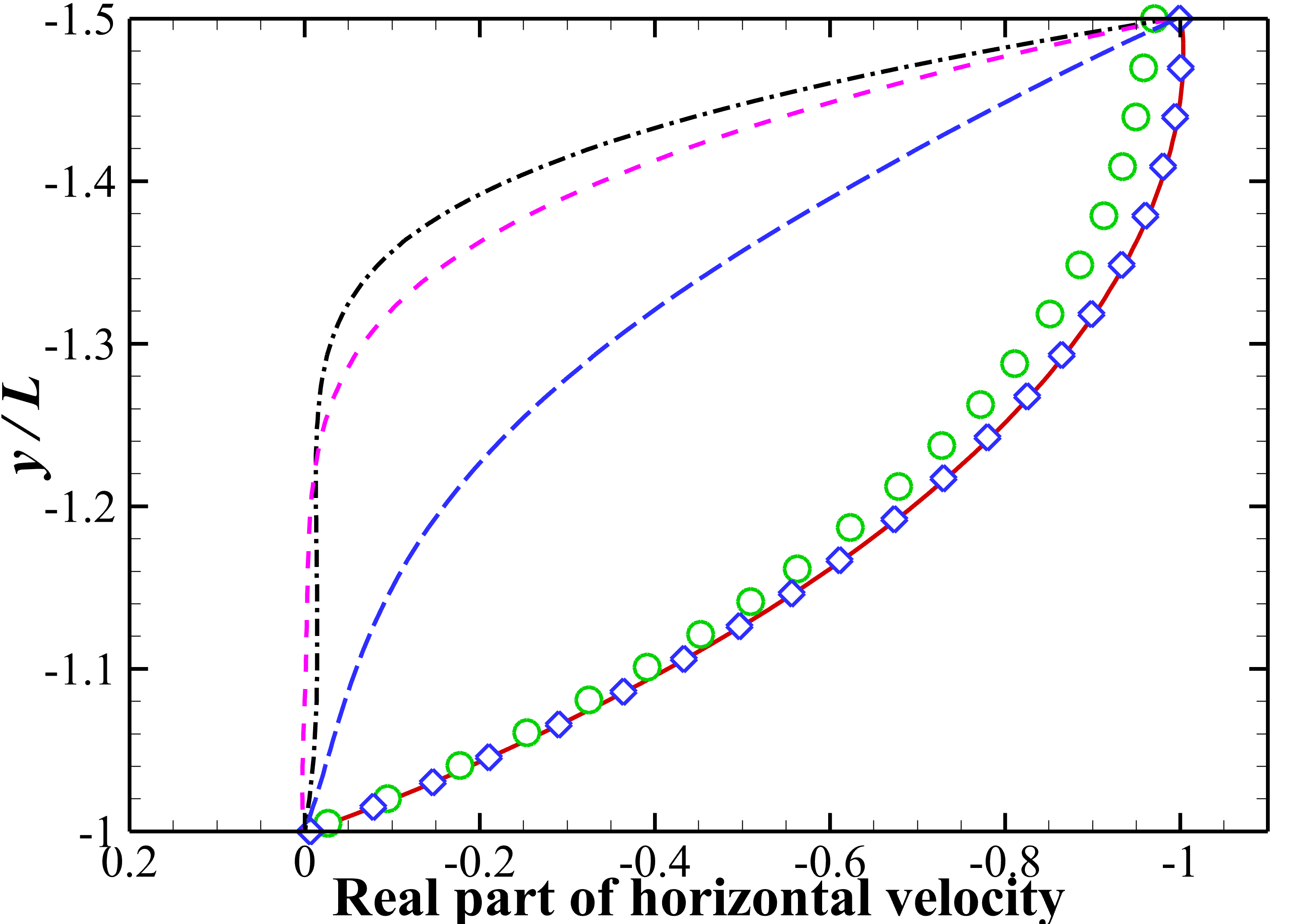}}
    \caption{The centerline streamwise velocity profiles at different iteration steps, when \(\delta_{rp}=1000\) and \(S=0.001\).  The reference NSF solution is computed on the same mesh.}
    \label{fig:cylinder-AP-ux}
\end{figure}

Figure~\ref{fig:cylinder-AP-ux} shows the evolution of the centerline velocity profiles for \((\delta_{rp},S)=(1000,0.001)\).
For CIS, the response remains confined to a thin, locally driven layer near the oscillating outer cylinder over thousands of iterations; even after \(10^4\) iterations, the velocity near the inner cylinder is still markedly underestimated.
In contrast, due to the super-convergence property of GSIS given in Eq.~\eqref{GSIS_super}, it approaches the reference profile  only after 5 iterations, and the curves become nearly indistinguishable after roughly 10 iterations.
This rapid global adjustment stems from the macroscopic synthetic equations, which efficiently propagate the boundary disturbances across the entire gap. 

Theoretically, according to Eqs.~\eqref{CIS_false} and~\eqref{GSIS_super}, to achieve the same level of convergence, the number of iterations required by the CIS would be $\delta_{rp}^4$ times greater than that required by the GSIS. Therefore, when \((\delta_{rp},S)=(1000,0.001)\), the large discrepancy between the CIS and the NSF reference solution in Fig.~\ref{fig:cylinder-AP-ux} can be attributed to two factors. First, after 37,570 iterations, although the error~\eqref{eq:residual} is less than $10^{-5}$, the solution has not fully converged due to false convergence. Second, since the CIS does not possess the AP property when the largest cell size is approximately 100 times larger than the mean free path, its solution exhibits substantial numerical dissipation.

Table~\ref{tab:cylinder-tab} further compares the convergence performance of CIS and GSIS.
In the near-continuum and low-frequency regime, CIS exhibits a severely deteriorated convergence rate and can become prohibitively slow; for instance, at $\delta_{rp}=1000$ and $S=0.001$, CIS requires more than $10^4$ iterations to reach the prescribed tolerance.
By contrast,  GSIS typically converges within $20\sim30$ iterations over the parameter space considered.
Although a single GSIS iteration is more expensive than CIS due to the need to solve additional macroscopic synthetic equations, GSIS attains a substantially shorter wall-clock time than CIS over a wide range of $(\delta_{rp}, S)$, with the advantage becoming particularly pronounced for large $\delta_{rp}$ and small $S$.

\begin{table}[h]
\centering
\caption{Iteration steps and CPU time to reach convergence for the oscillatory shear-driven flow between two eccentric cylinders. The code is implemented in double precision with OpenMP parallelization and executed on AMD EPYC 7763 processor (2.45GHz) using 12 threads.
}
\label{tab:cylinder-tab}
\setlength{\tabcolsep}{4.5pt}
\renewcommand{\arraystretch}{1.05}
\begin{tabular}{cccccccc}
\toprule
\multirow{2}{*}{$\delta_{rp}$} &
\multirow{2}{*}{$S$} &
\multirow{2}{*}{$N_{\text{cell}}$} &
\multicolumn{2}{c}{Iteration steps} &
\multicolumn{2}{c}{Total CPU time (s)} \\
\cmidrule(lr){4-5}\cmidrule(lr){6-7}
 &   &  & CIS & GSIS & CIS & GSIS \\
\midrule
1    & 1.0   & 6,000   & 64    & 24 & 20      & 14  \\
10   & 1.0   & 6,000   & 414   & 25 & 134     & 14 \\
100  & 0.1   & 12,000  & 10,047 & 29 & 5,349  & 25 \\
1000 & 0.001 & 12,000  & 37,570 & 27 & 23,787 & 32 \\
\bottomrule
\end{tabular}
\end{table}

\begin{figure}[t!]
    \centering
    \subfigure
    {\includegraphics[width=0.32\linewidth]
    {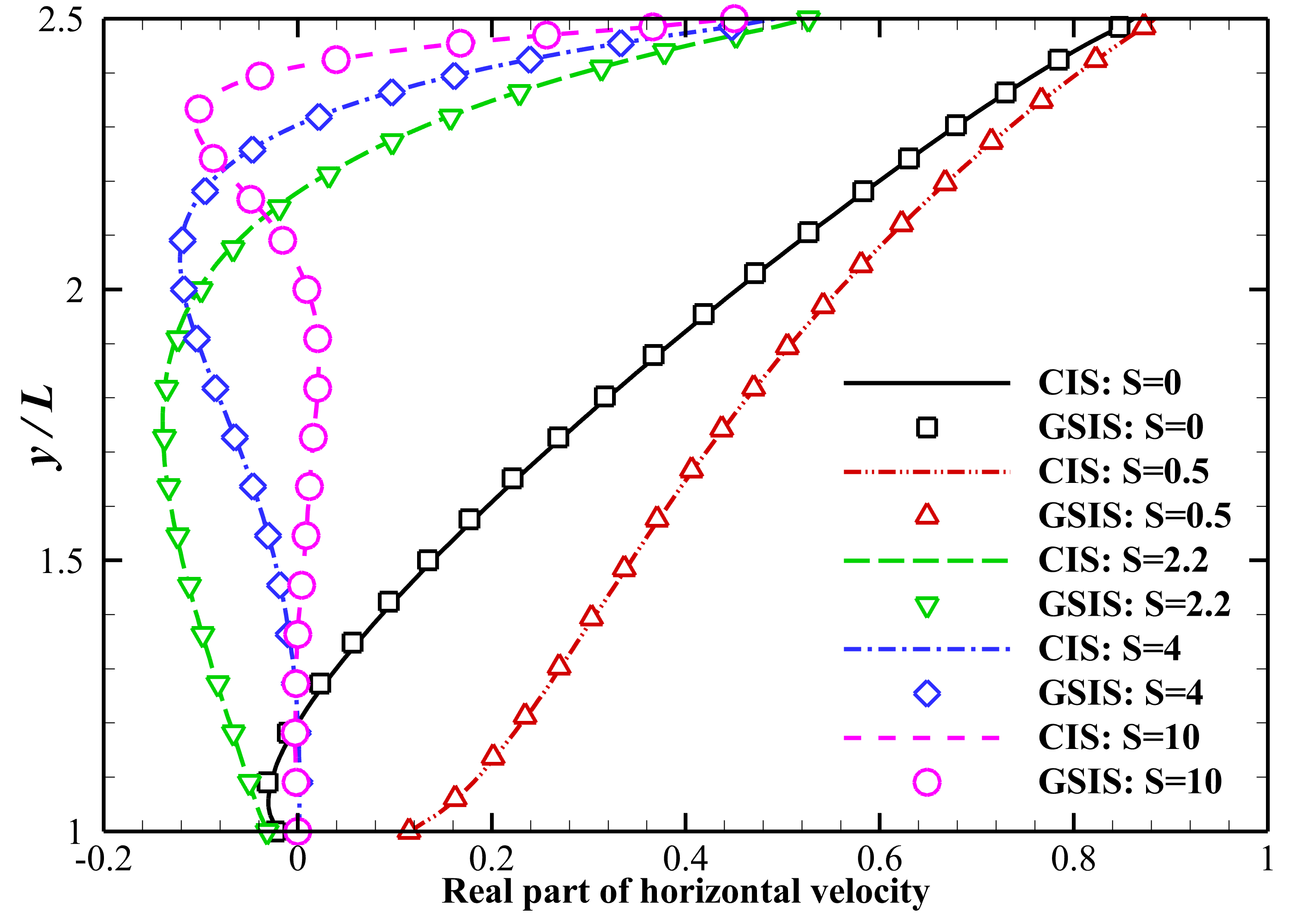}}
    {\includegraphics[width=0.32\linewidth]
    {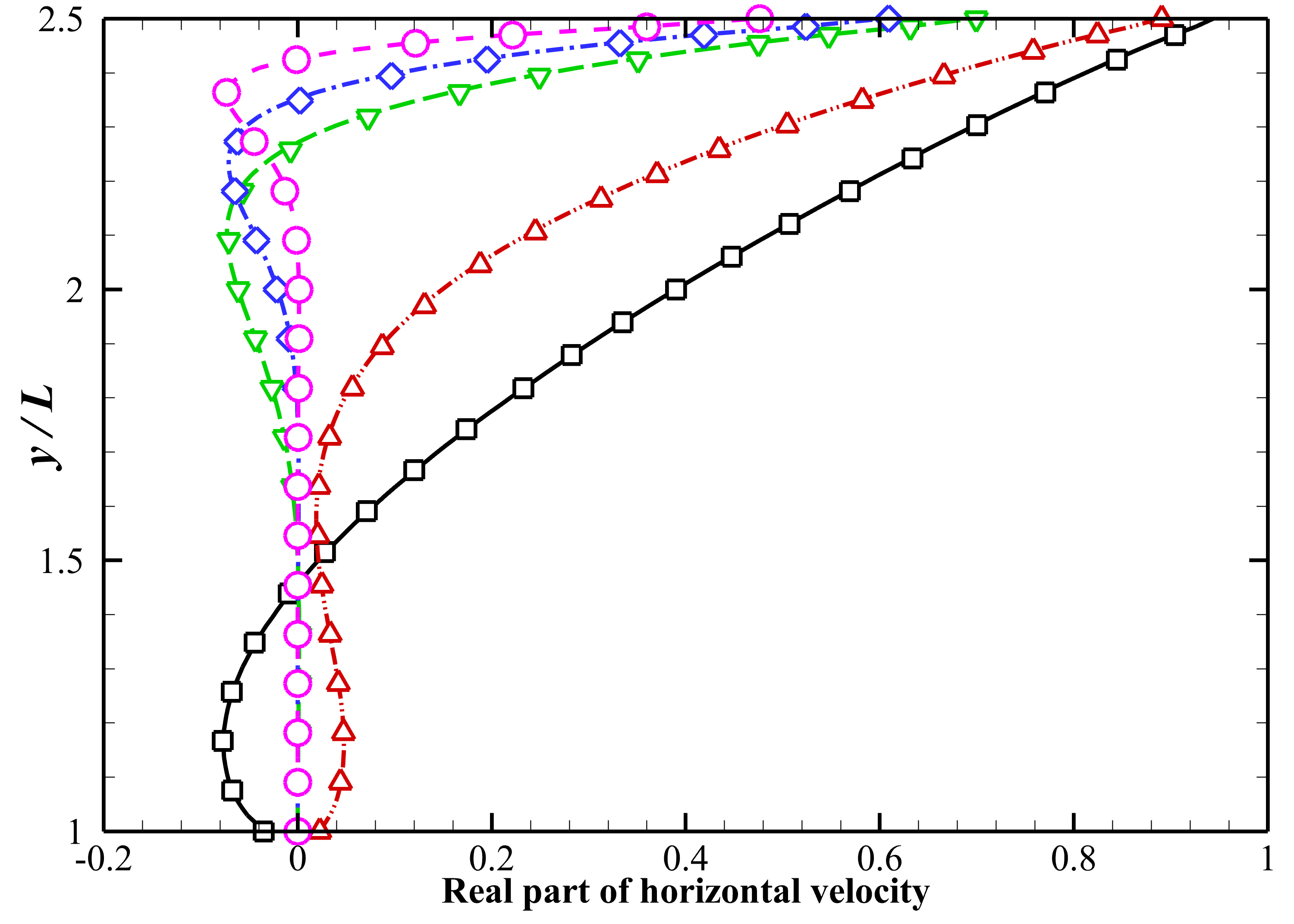}}
    {\includegraphics[width=0.32\linewidth]
    {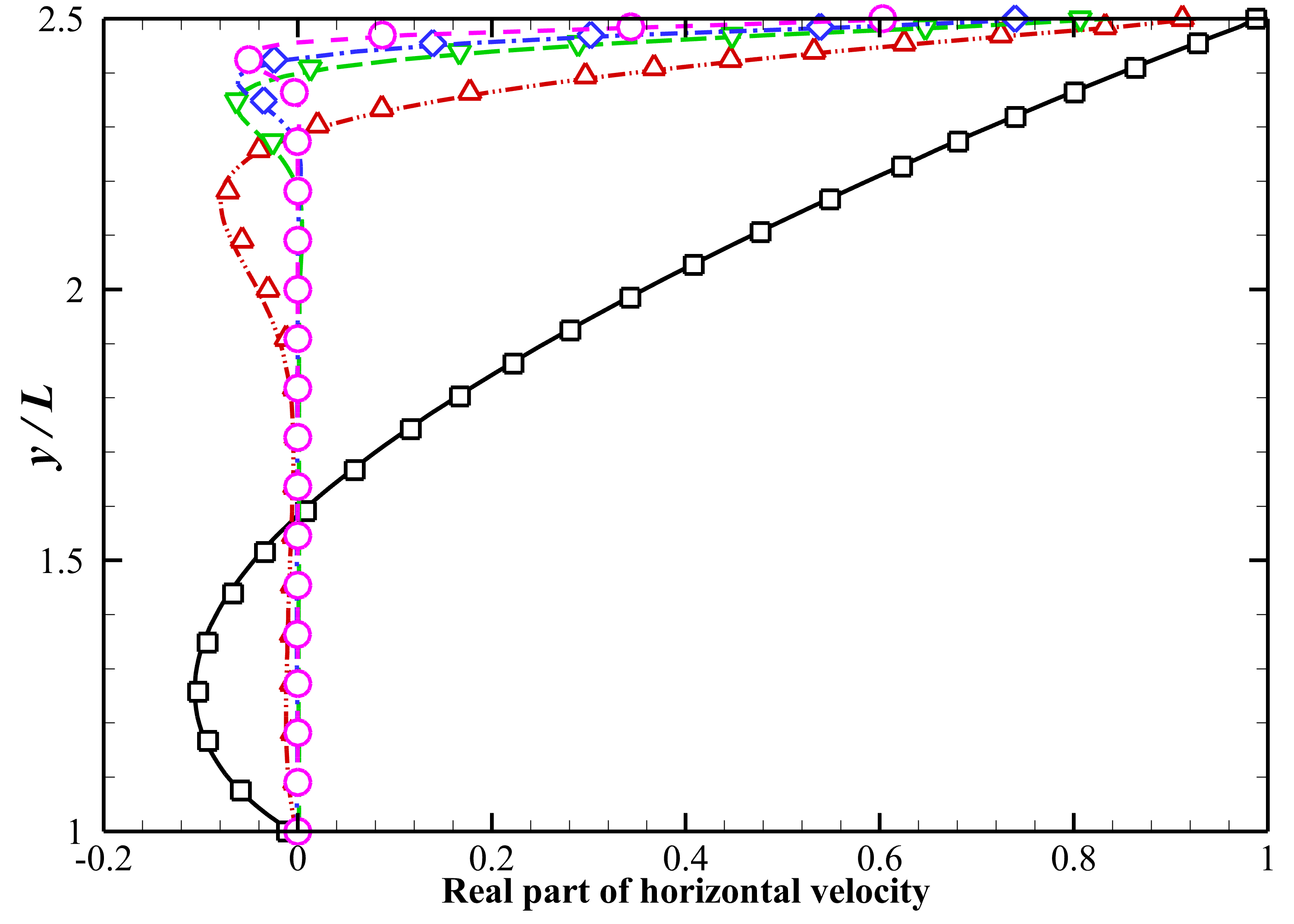}}\\
    {\includegraphics[width=0.32\linewidth]
    {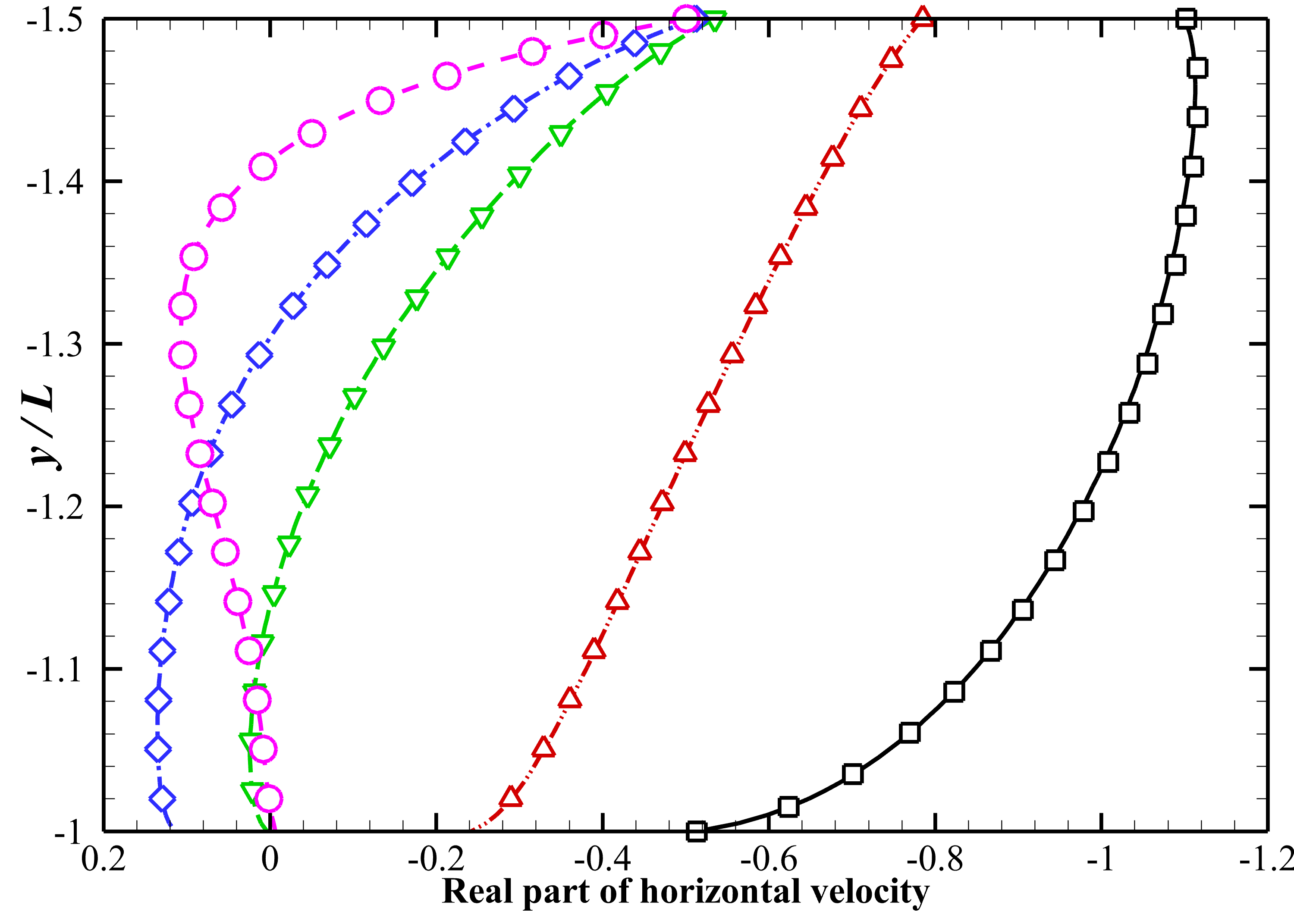}}
    {\includegraphics[width=0.32\linewidth]
    {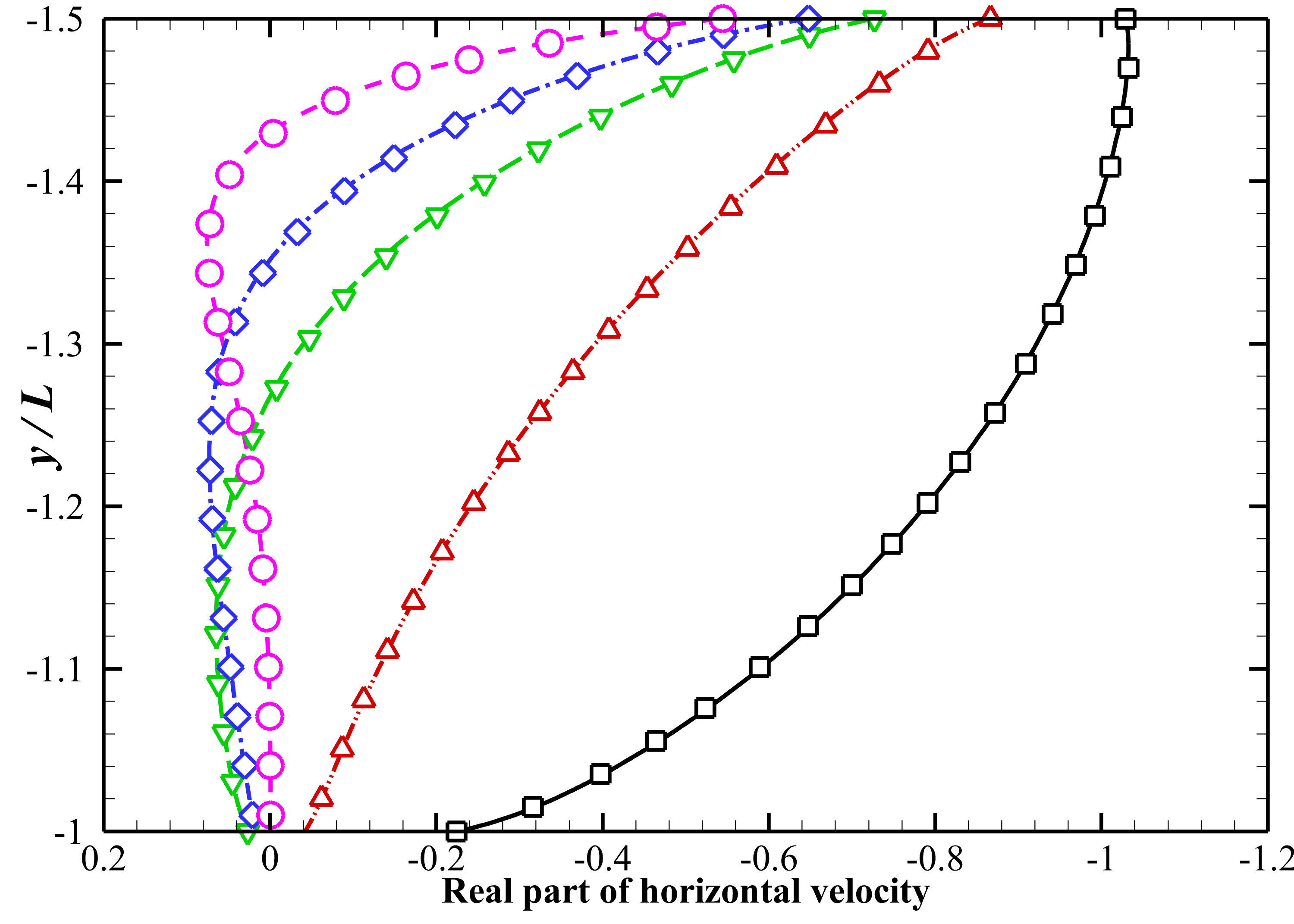}}
    {\includegraphics[width=0.32\linewidth]
    {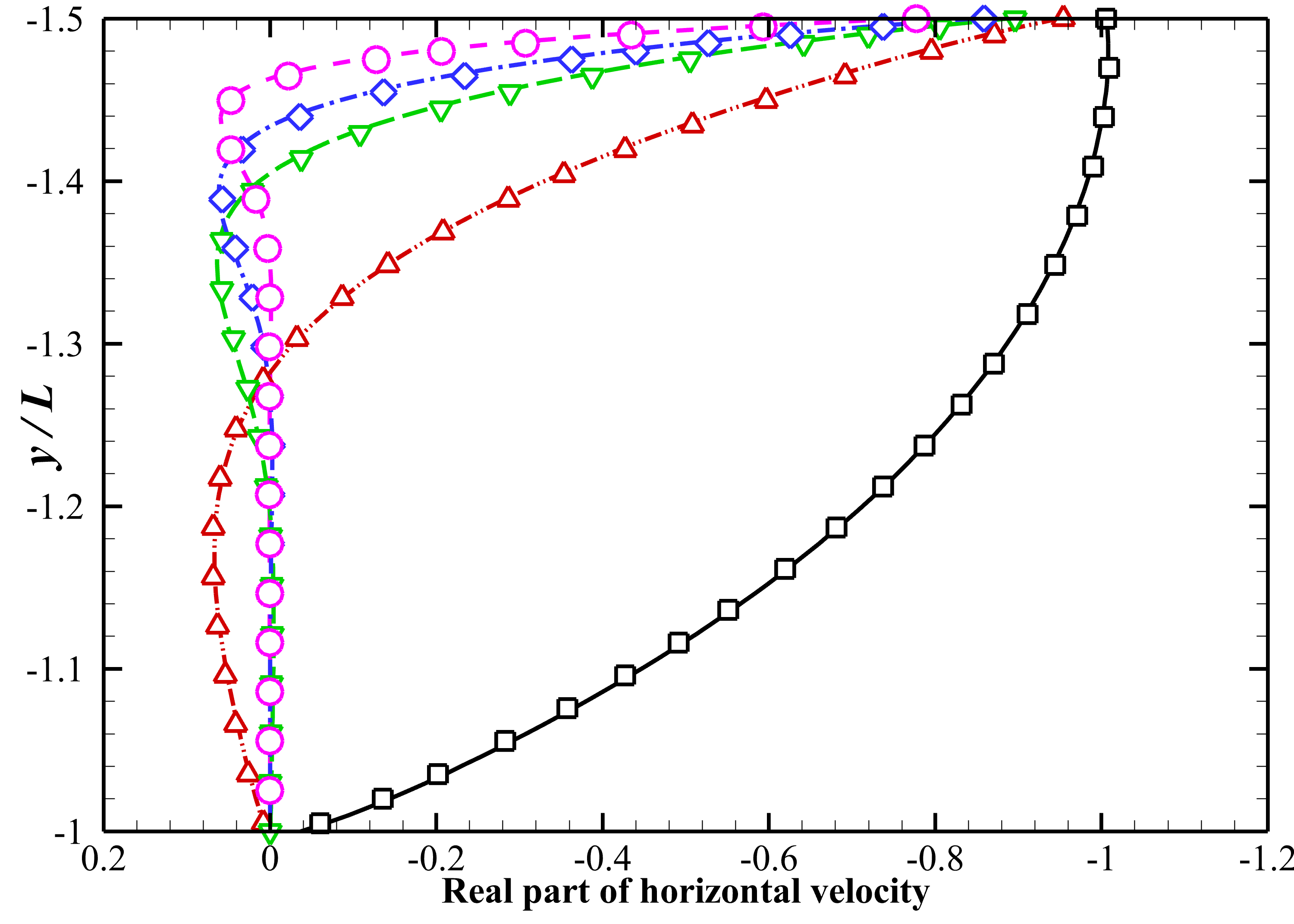}}
    \caption{The real part of the horizontal velocity at \(x=0\). From left to right: \(\delta_{rp}=1,10,100\). The top and bottom rows show the velocity profiles above and below the inner cylinder, respectively. 
    }
    \label{fig:cylinder-u_real}
\end{figure}

\subsubsection{Variation of shear-force with $\delta_{rp}$ and $S$}

To quantify the oscillatory drag on the outer cylinder, we define the shear force as
\begin{equation}
\gamma=-\int_{\Gamma_o} \bm{t}_w\cdot\left(\bm{\Pi}_w\cdot \bm{n}\right)\,dS,
\label{eq:gamma_def}
\end{equation}
where $\Gamma_o$ is the outer-cylinder surface.

To relate the shear response to the flow structure, we introduce the classical viscous penetration depth of oscillatory Stokes flows~\cite{schlichting2016boundary}.
It represents the e-folding decay length of the oscillatory velocity amplitude away from an oscillating wall.
In the NSF limit, the corresponding non-dimensional penetration depth scales as $\delta_s \propto (\delta_{rp}S)^{-\frac{1}{2}}$.
Although the present configuration is annular and may be rarefied, 
$\delta_s$ provides a useful reference scale: a smaller $\delta_s$ implies a more localized oscillatory motion near the moving wall and a steeper near-wall gradient, hence a larger wall shear.

\begin{figure}[h]
    \centering
    \includegraphics[width=0.45\linewidth]
    {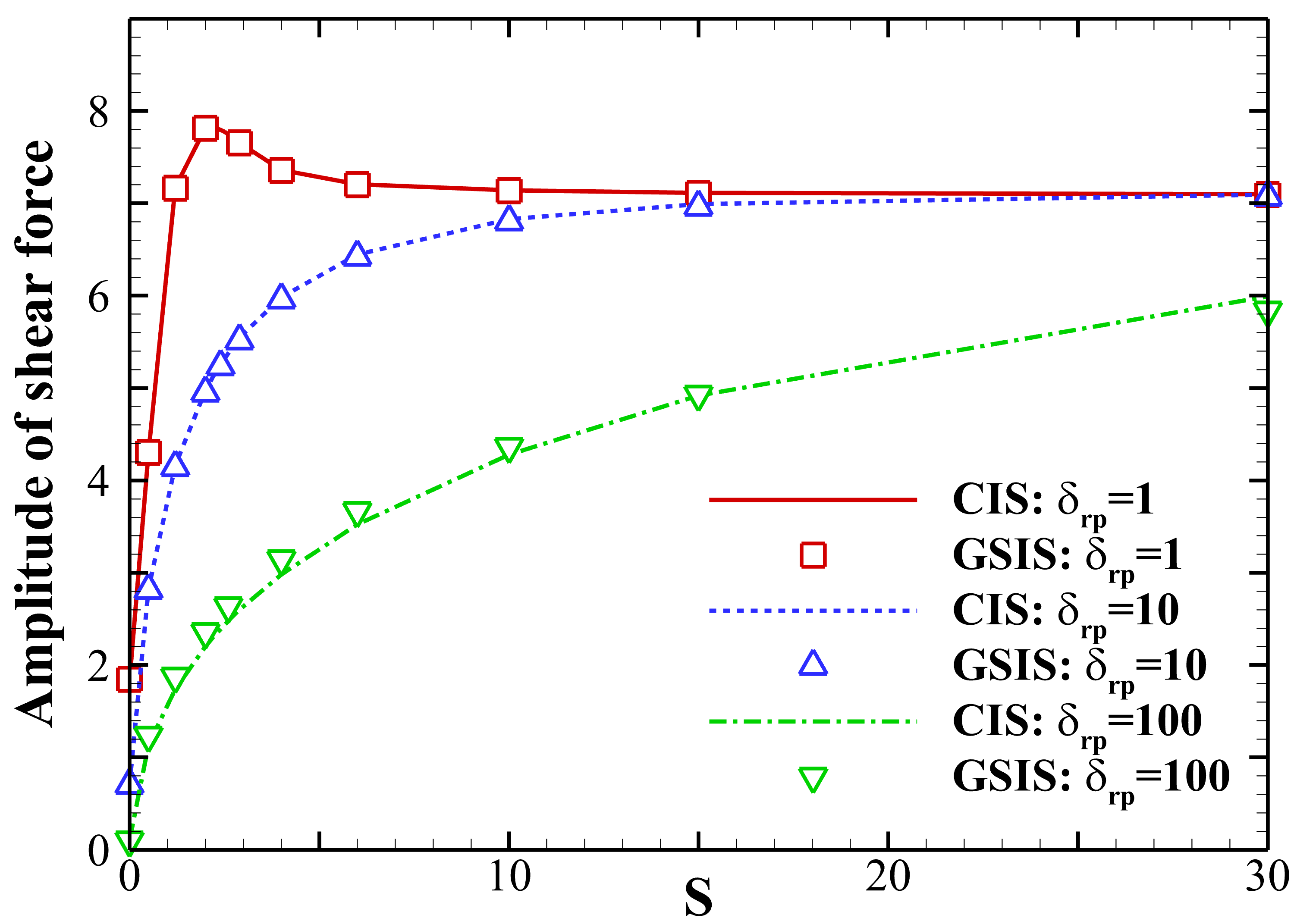}
    \hspace{0.25cm}
    \includegraphics[width=0.45\linewidth]
    {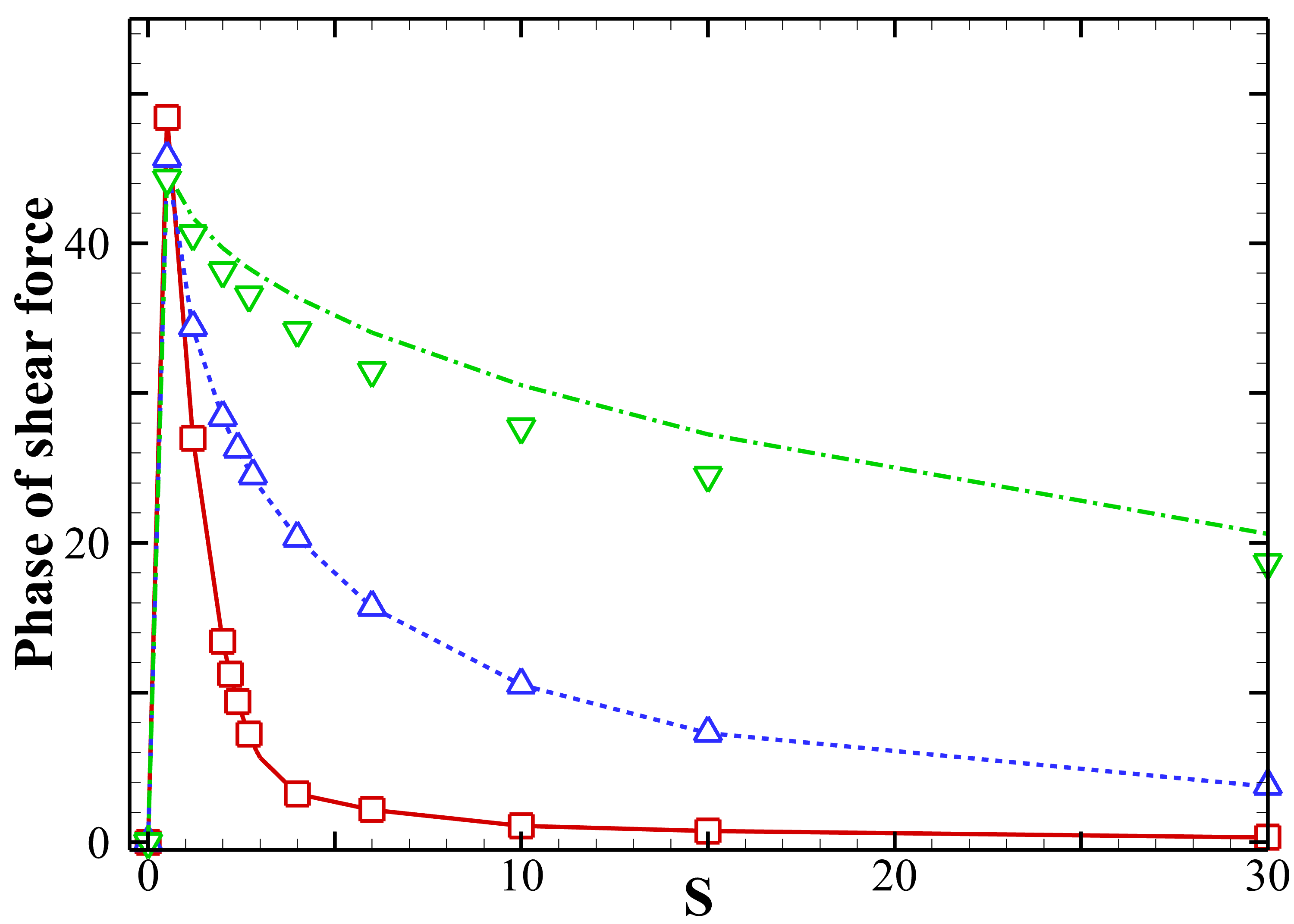}
    \caption{
    Amplitude and phase of shear force \(\gamma\) versus Strouhal number \(S\) for \(\delta_{rp}=1,10,100\). 
    }
    \label{fig:cylinder-shear}
\end{figure}

This behavior is reflected by the streamwise-velocity profiles in Fig.~\ref{fig:cylinder-u_real}.
With increasing $\delta_{rp}$ and $S$, the near-wall gradients become steeper (for both the wider and narrower gaps), consistent with a reduced effective penetration depth.
Accordingly, Fig.~\ref{fig:cylinder-shear} shows that the amplitude $|\gamma|$ generally increases with $S$ and approaches a high-frequency plateau; once the response is confined to a thin near-wall region, further increases in $S$ no longer enhance momentum penetration across the gap, and the shear tends toward a finite high-frequency limit~\cite{wu2014oscillatory}.



Besides the overall saturation trend, an overshoot is observed for $\delta_{rp}=1$ around $S\approx 2.2$ (i.e., $Kn_t=O(1)$), which marks a transition regime where the relaxation time becomes comparable to the oscillation period.
In this regime, the continued reduction of the effective penetration depth further steepens the near-wall gradient, while finite-relaxation non-equilibrium effects become appreciable, so that the shear can transiently exceed its high-frequency plateau.
For $\delta_{rp}=10$ and $100$, $Kn_t$ is smaller over the same range of $S$ and the transition is shifted to higher frequencies; consequently, the shear increases more monotonically and no pronounced overshoot is observed.

The corresponding phase response is consistent with this picture.
The phase rises rapidly at low $S$ and then decays toward $0^\circ$ as $S$ increases, indicating that the shear force becomes nearly in phase with the wall velocity in the high-frequency limit.
For a fixed $S$, decreasing $\delta_{rp}$ increases $Kn_t$, bringing the response closer to the high-frequency regime and thus accelerating the decay of the phase toward zero.


\subsection{Squeeze-film damping }

\begin{figure}[t]
    \centering
    \subfigure[]
    {\includegraphics[width=0.5\linewidth]
    {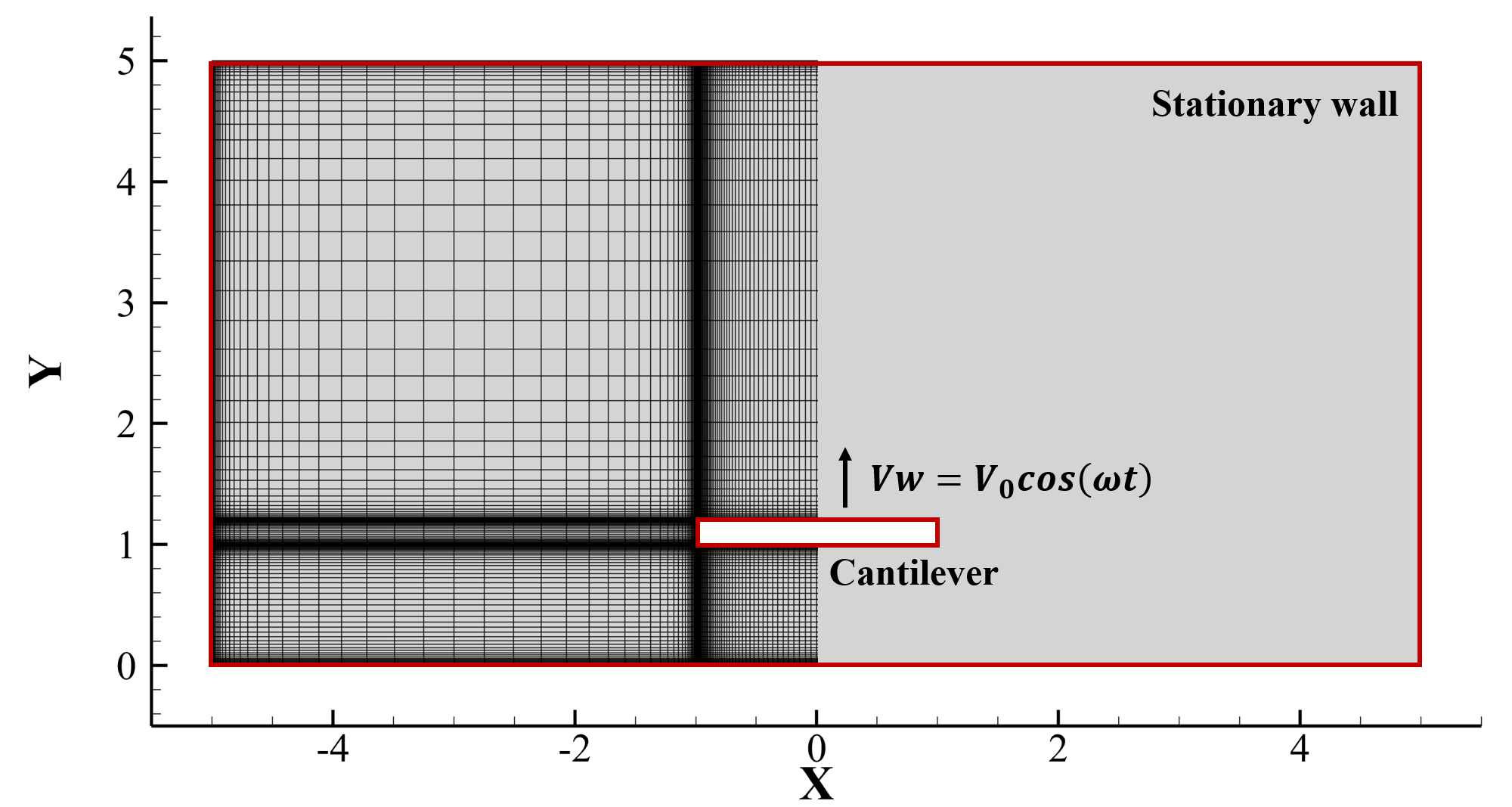}}\\
    \subfigure[]
    {\includegraphics[width=0.48\linewidth]
    {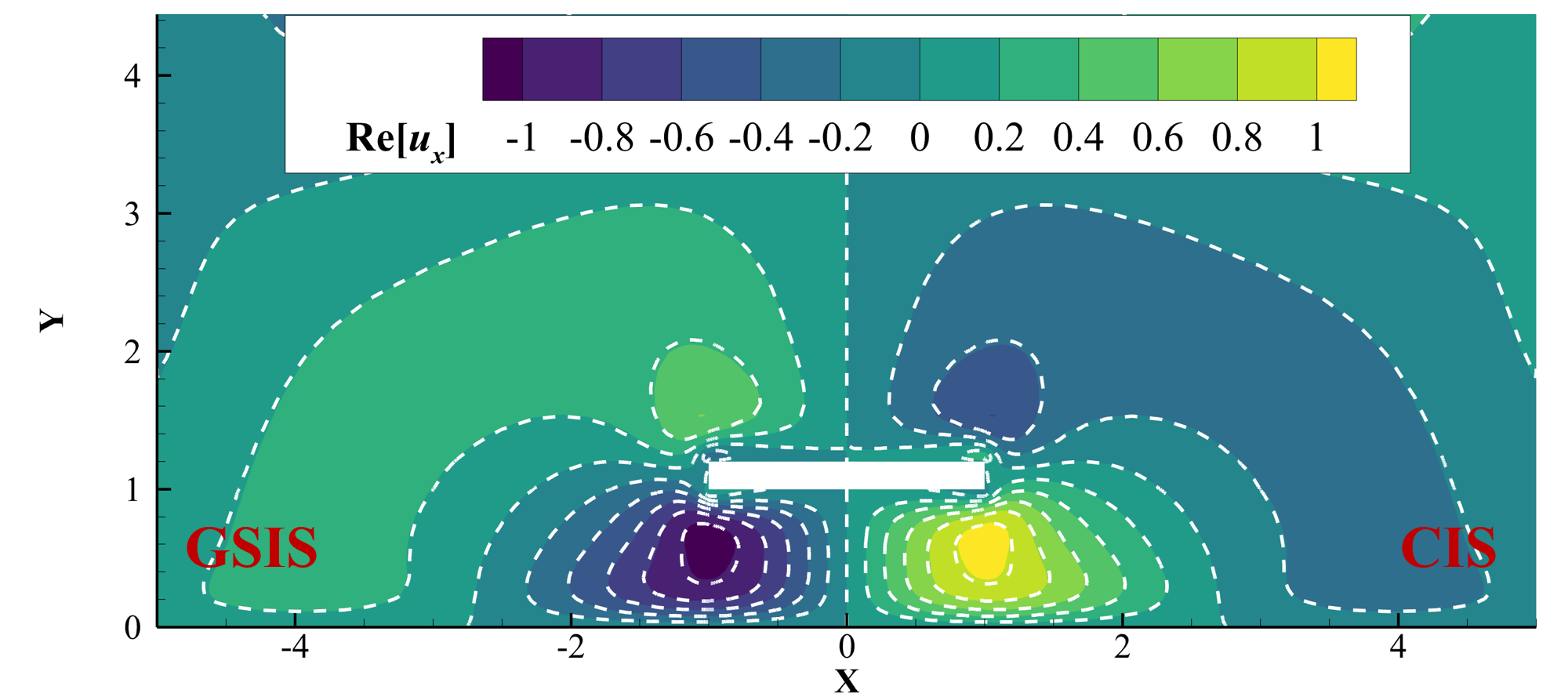}}
    \subfigure[]
    {\includegraphics[width=0.48\linewidth]
    {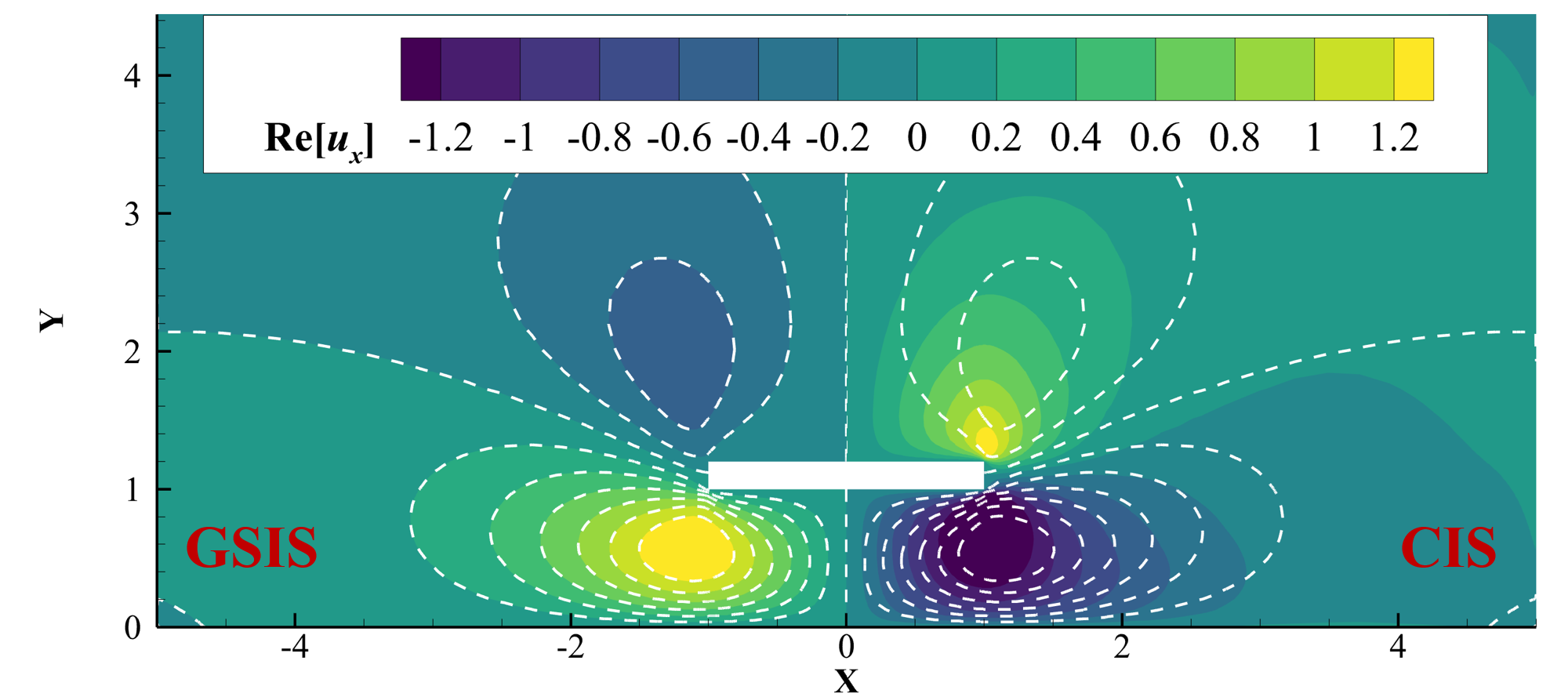}}
    \caption{Assessment of GSIS's AP property for the oscillating cantilever configuration. 
    (a) Geometry and computational mesh. 
    (b) Contours of the real part of the horizontal velocity, \(\Re(u_x)\) for \(\delta_{rp}=50,S=1\). 
    (c) Contours of \(\Re(u_x)\) for \(\delta_{rp}=1000,S=0.001\). 
    The white dashed lines denote the reference solution used for comparison. The horizontal flow velocity is asymmetric about the line $x=0$, and the GSIS and CIS results are shown in the left and right half regions, respectively.  
    }
    \label{fig:cantilever}
\end{figure}

We consider an oscillating cantilever positioned above a stationary substrate with a gap height \(L\), which is adopted as the reference length. The geometry and mesh are shown in Fig.~\ref{fig:cantilever}. The computational domain is bounded by stationary walls on the outer boundary, while a rigid cantilever is embedded inside the domain. The lower horizontal wall represents the stationary substrate, and the cantilever is located above it, forming a narrow gap of height \(L\) where the squeeze-film flow develops. All outer boundaries are treated as stationary walls, whereas the cantilever surface is a moving wall that undergoes a small-amplitude harmonic vibration in the vertical direction,
$
V_w(t)=V_0\cos(\omega t)$.
The spatial grid contains \(N_{\text{cell}}=28{,}000\) cells, with local refinement near solid boundaries down to a minimum spacing of \(0.001L\) to resolve near-wall variations. Since the vibration-induced velocity is much smaller than the molecular thermal speed, the governing equations can be linearized and solved directly in the frequency domain, avoiding long time-marching to reach the periodic steady state. For this case, \(\xi = V_0/v_m\).

All solid boundaries are modeled with fully diffuse reflection. The cantilever surface is driven by a spatially uniform oscillatory velocity boundary condition, while the other boundaries are treated as stationary walls. The reflected distribution function on the cantilever surface is given by 
\begin{equation}
    h=\frac{2}{\pi}\int_{v_n>0} v_n \exp(-v^2)\, h \, d\bm{v}+2 v_y
    -\sqrt{\pi}\, n_y,
  \quad \text{for}~  v_n<0, 
\end{equation}
where $n_y=\bm{n}\cdot\bm{e}_y$ is the $y$-component of the outward unit normal vector $\bm{n}$ on the cantilever surface. 
For stationary walls, the boundary condition takes the same form but without the $2v_y$ and $-\sqrt{\pi}\,n_y$ terms.

In the molecular velocity space, a non-uniform discrete velocity grid is employed with \(32\) points per velocity direction over the range \([-6,6]\), and the node distribution follows Eq.~\eqref{eq:velocity_grid}.

\subsubsection{Asymptotic-preserving and super-convergence}

To validate the accuracy of GSIS across regimes, we examine two representative cases, \((\delta_{rp},S)=(50,1)\) and \((1000,0.001)\), as shown in Fig.~\ref{fig:cantilever}. For \((\delta_{rp},S)=(50,1)\), the CIS solution on the same mesh is used as the reference, and GSIS agrees well with CIS. For the near-continuum, low-frequency case of \((\delta_{rp},S)=(1000,0.001)\), rarefaction effects are negligible and the solution should approach the NSF limit; thus, the NSF solution on the same mesh is taken as the reference. In this case, GSIS matches the reference closely, whereas the CIS result (shown after \(3\times 10^4\) iteration, with a residual level of \(10^{-4}\)) still exhibits a substantial deviation from the NSF reference.

Table~\ref{tab:cantilever-tab} compares the iteration counts and total CPU time for CIS and GSIS from rarefied to near-continuum regimes. The number of CIS iterations increases dramatically with \(\delta_{rp}\), while the GSIS iteration count remains nearly insensitive to the regime and typically converges within 20--30 iterations. Consequently, GSIS yields a substantial reduction in time-to-solution.

\begin{table}[t]
\centering
\caption{Iteration counts and total CPU time to reach the convergence criterion \eqref{eq:residual} for the oscillating cantilever flow.
All cases use the same spatial and velocity discretizations, with $N_{\text{cell}}=2.8\times 10^{4}$ and $N_v=1024$.
The code is implemented in double precision with OpenMP and executed on an AMD EPYC 7763 (2.45\,GHz) using 12 threads.
For $(\delta_{rp},S)=(1000,0.001)$, CIS does not reach the tolerance within $10^4$ iterations; hence the corresponding entries are omitted.}
\label{tab:cantilever-tab}
\setlength{\tabcolsep}{6pt}
\renewcommand{\arraystretch}{1.05}
\begin{tabular}{cccccc}
\toprule
\multirow{2}{*}{$\delta_{rp}$} & 
\multirow{2}{*}{$S$} & 
\multicolumn{2}{c}{Iteration steps} & 
\multicolumn{2}{c}{Total CPU time (s)} \\
\cmidrule(lr){3-4}\cmidrule(lr){5-6}
& & CIS & GSIS & CIS & GSIS \\
\midrule
1    & 1.0   & 58     & 27 & 67     & 66 \\
10   & 1.0   & 1,091  & 29 & 1,209  & 67 \\
100  & 1.0   & 10,725 & 30 & 11,765 & 80 \\
1000 & 0.001 & ---    & 27 & ---    & 60 \\
\bottomrule
\end{tabular}
\end{table}

\subsubsection{Frequency response}
We quantify the SFD response by analyzing the magnitude and phase of the vertical hydrodynamic force acting on the cantilever. Under the present non-dimensionalization, the stress tensor is expressed as
    $\bm{P}=p\bm{I} + \bm{\Pi}$,
where \(p=\rho+\tau\) is the pressure. The resultant force is obtained from the surface integral
\begin{equation}
    \bm{F}=-\int_{\Gamma_b} (\bm{P}\cdot \bm{n}) dS,
\end{equation}
with \(\Gamma_b\) denoting the cantilever surface. 

Owing to geometric symmetry, the horizontal force component is negligible, and we focus on the complex vertical load \(F_y\), which serves as the SFD response in what follows. The vertical load receives contributions from (i) the pressure-related term \(p\,dS_y\) and (ii) the deviatoric-stress terms \(\Pi_{xy}\,dS_x+\Pi_{yy}\,dS_y\). Accordingly, the magnitude and phase of the SFD are obtained from \(|F_y|\) and \(\arg(F_y)\), respectively.

\begin{figure}[t!]
\centering
    \includegraphics[width=0.4\linewidth]{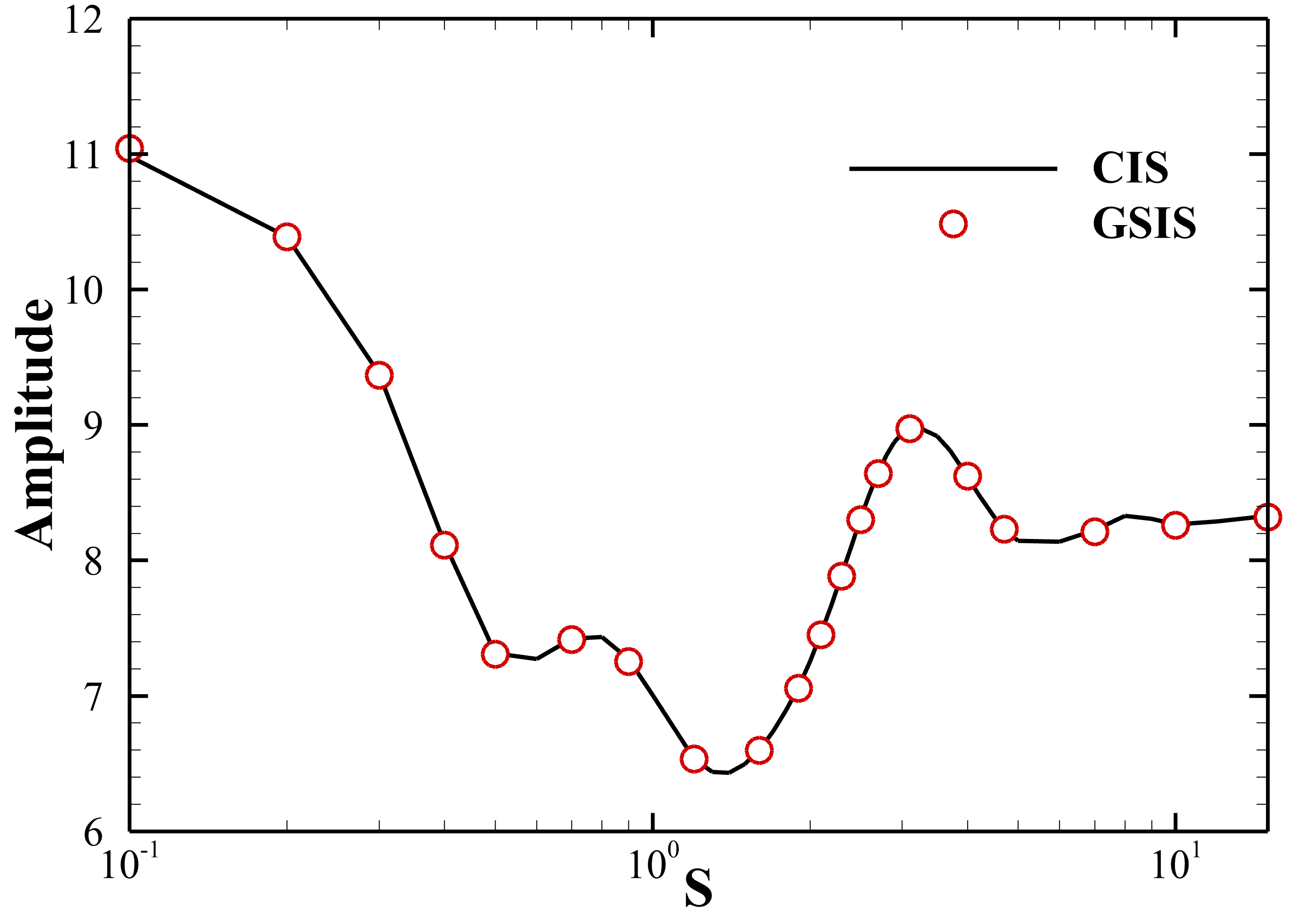}
    \includegraphics[width=0.4\linewidth]{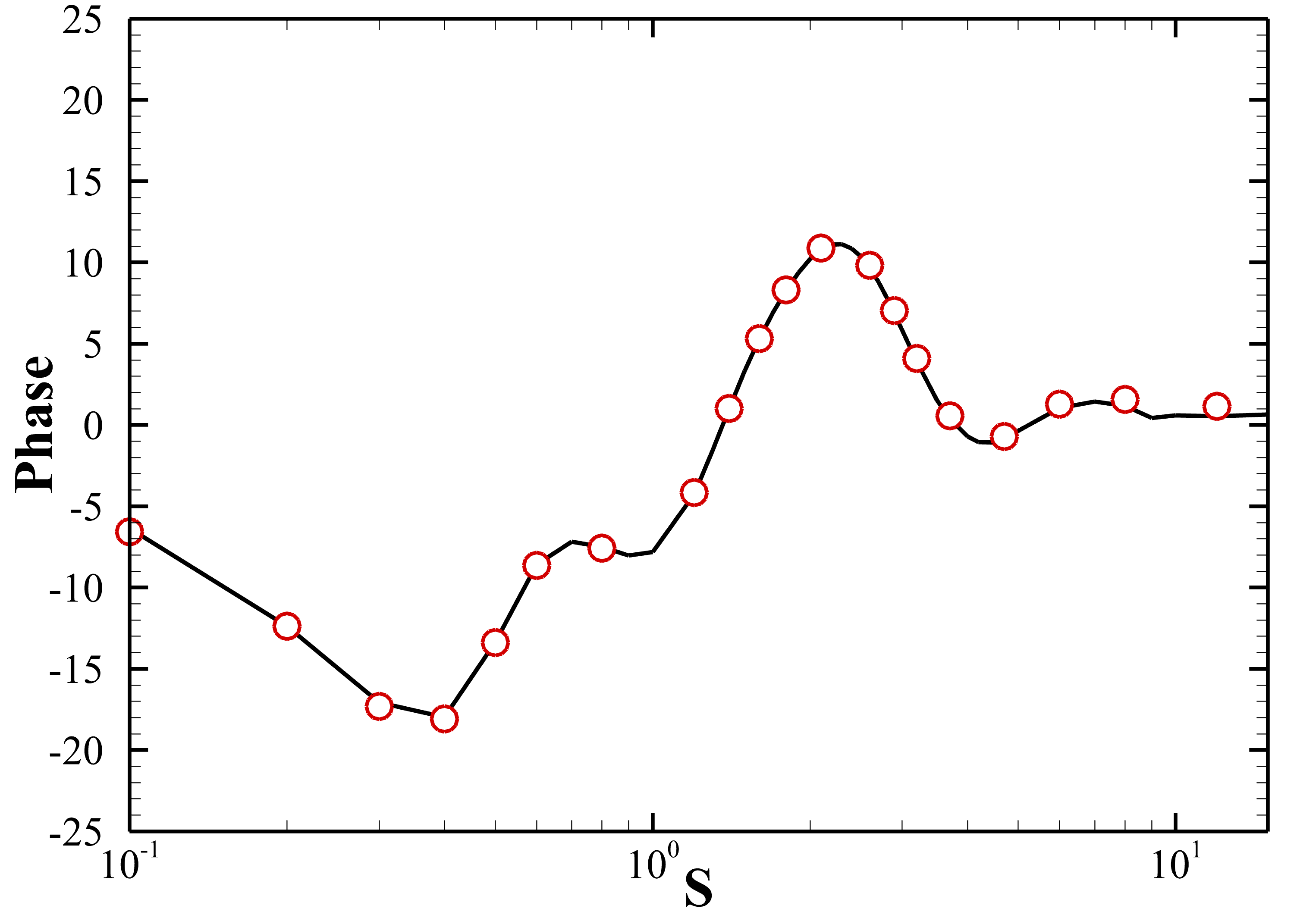}
    \includegraphics[width=0.4\linewidth]{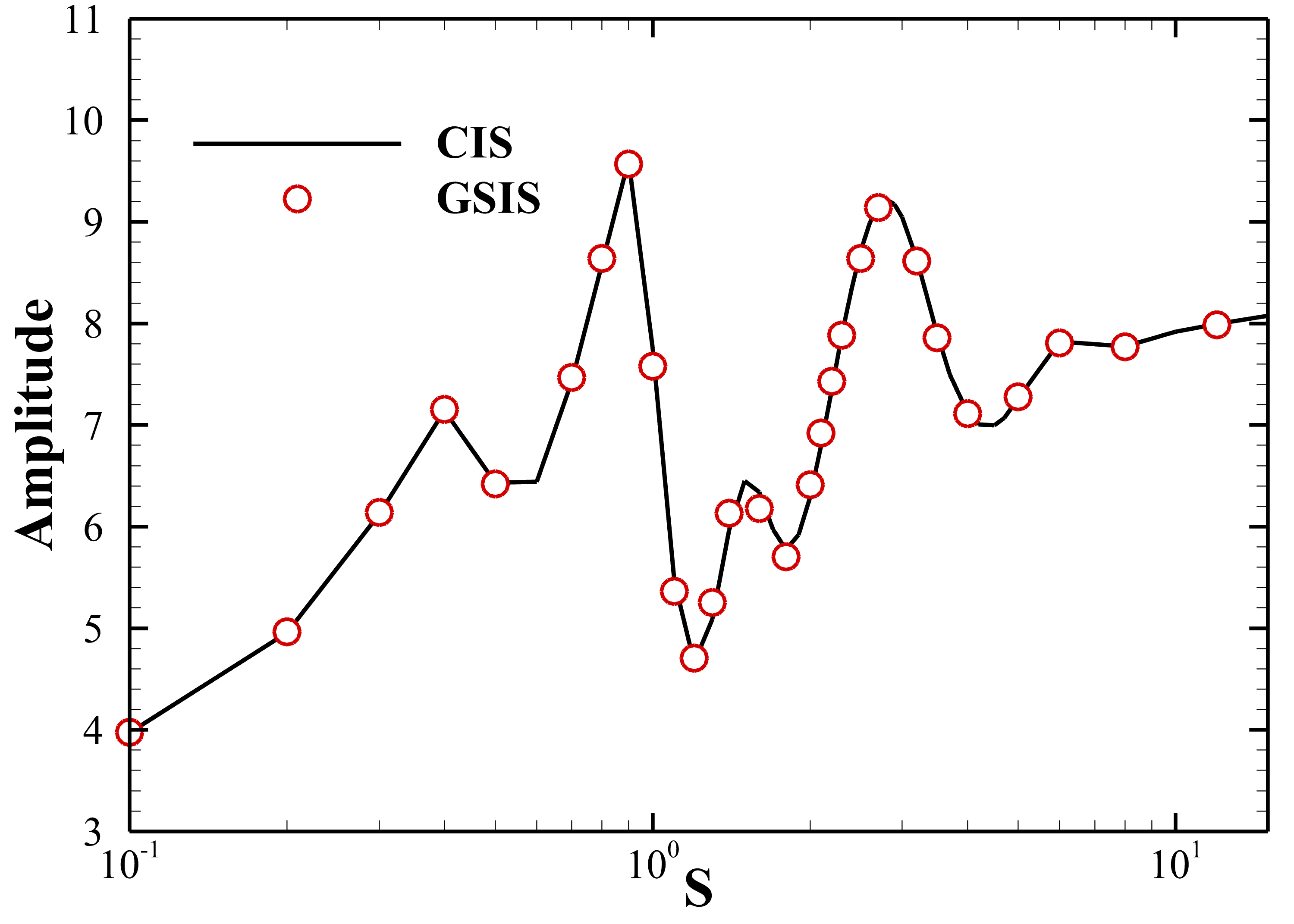}
    \includegraphics[width=0.4\linewidth]{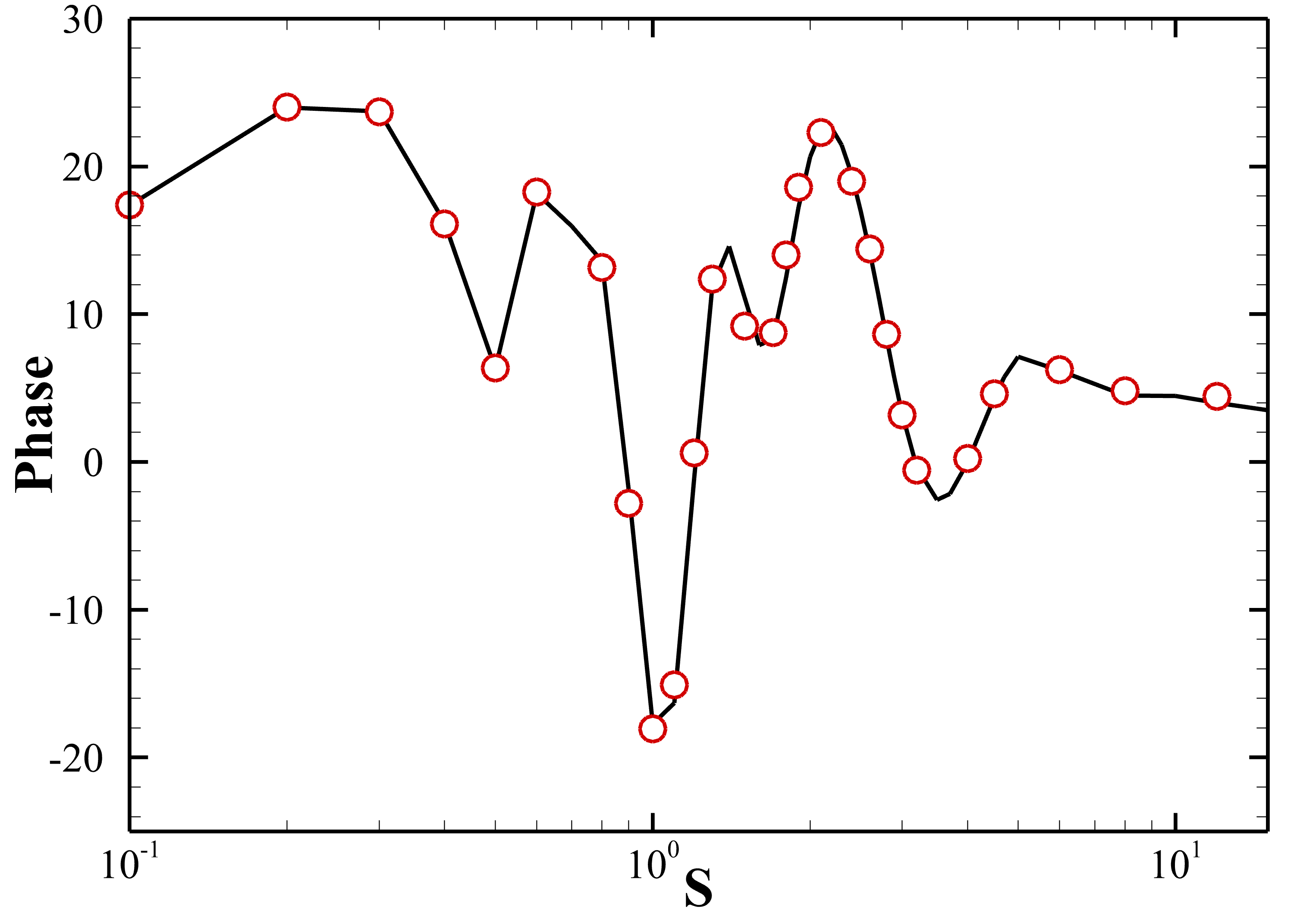}
    \includegraphics[width=0.4\linewidth]{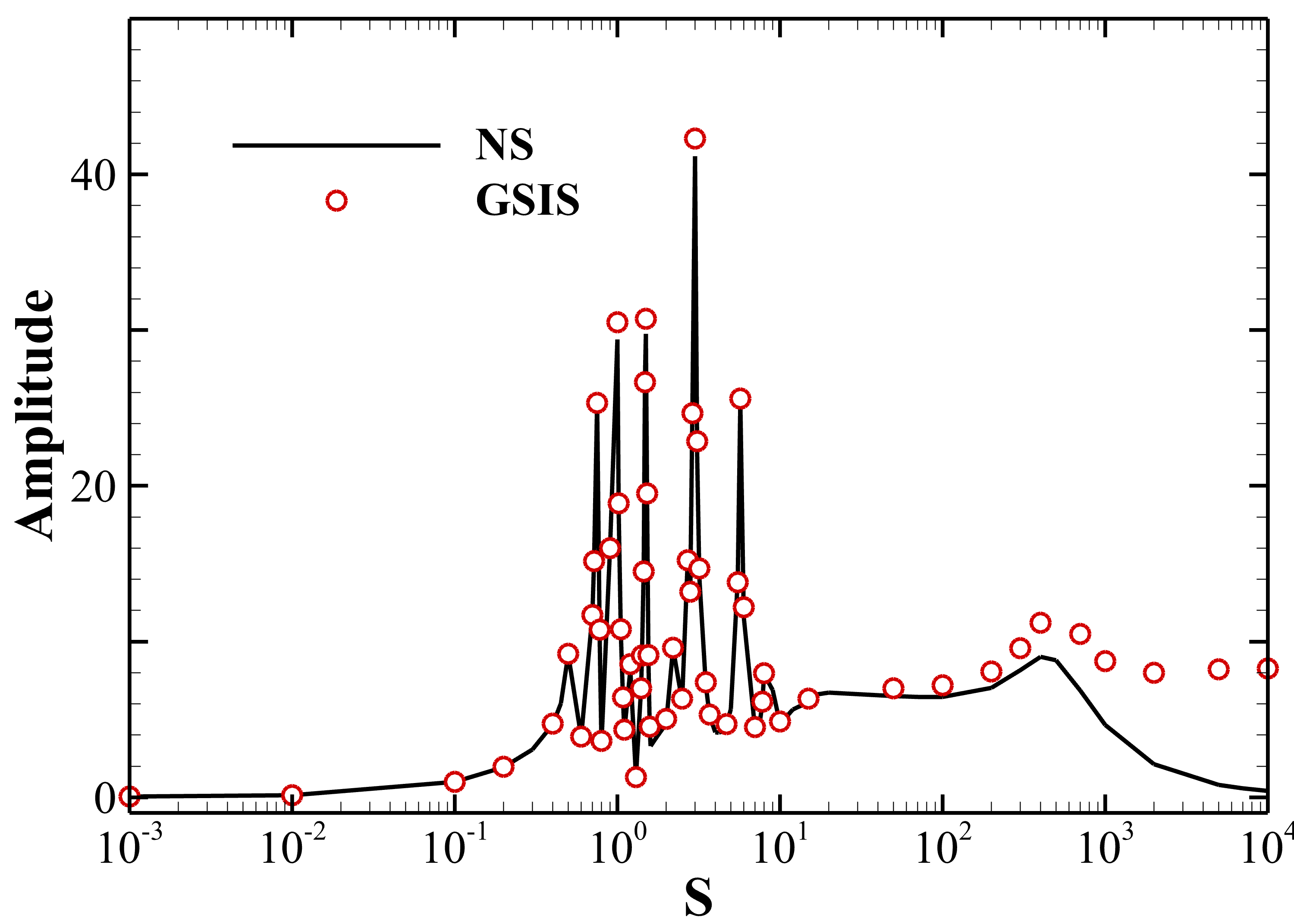}
    \includegraphics[width=0.4\linewidth]{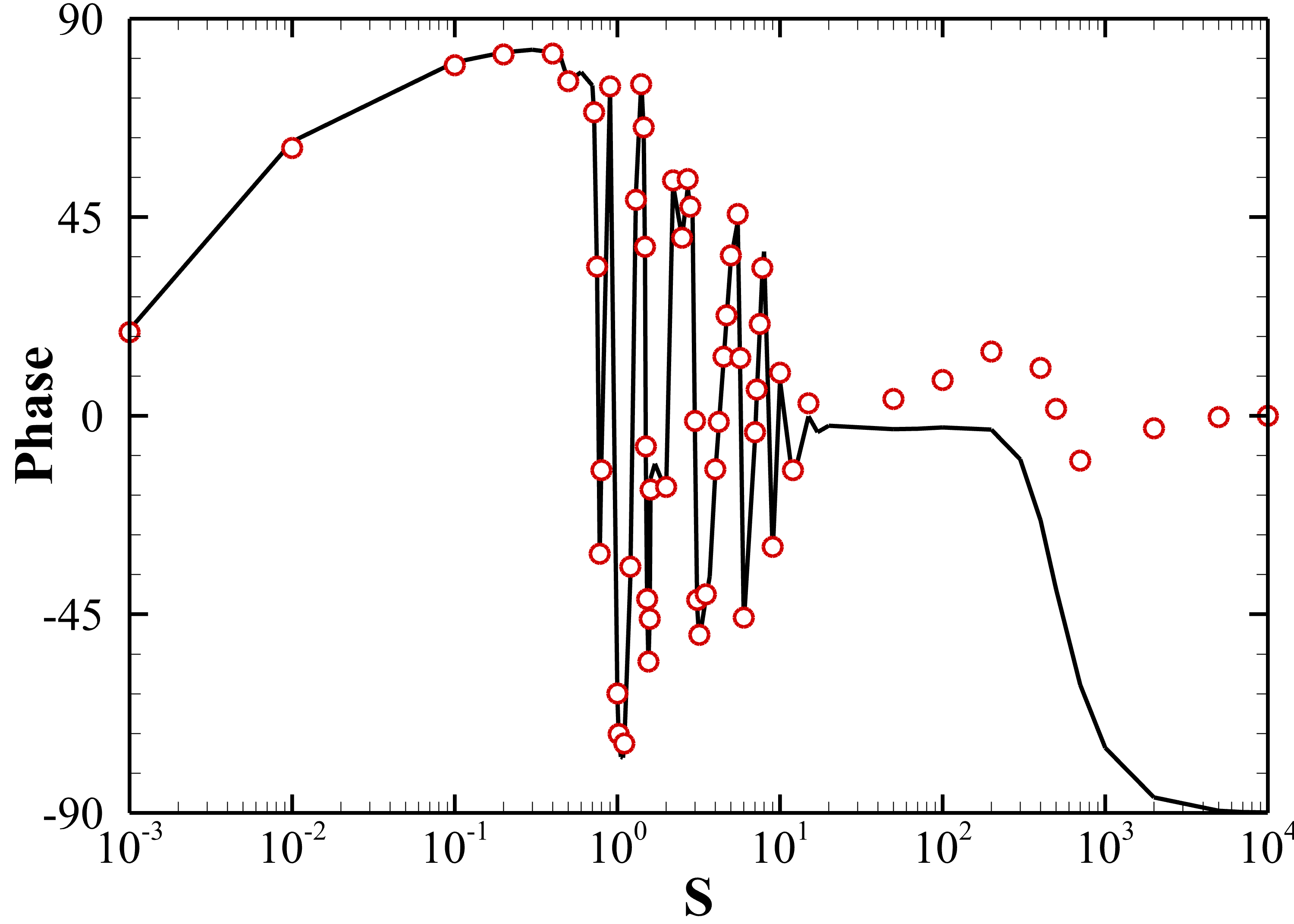}
    \caption{Magnitude and phase of the squeeze-film damping  as functions of the Strouhal number \(S\). From top to bottom: \(\delta_{rp}=1,\,10,\) and \(1000\).}
    \label{fig:cantilever-wall-mag-phase}
\end{figure}
Figure~\ref{fig:cantilever-wall-mag-phase} shows the magnitude and phase of the SFD as functions of the \(S\) for \(\delta_{rp}=1,\,10,\) and \(1000\). For \(\delta_{rp}=1\) and \(10\), the CIS and GSIS results are nearly indistinguishable over the entire range of \(S\) considered, indicating that GSIS preserves kinetic accuracy in rarefied regimes. For \(\delta_{rp}=1000\), results from both NSF and GSIS are reported.

For \(\delta_{rp}=1000\), GSIS agrees well with the NSF solution up to approximately \(S\approx 100\), for which the corresponding temporal Knudsen number is about \(Kn_t\sim 0.1\). When \(Kn_t\gtrsim 0.1\) (i.e., at higher frequencies), the NSF and GSIS trends begin to diverge, reflecting the breakdown of the continuum-in-time assumption. In the continuum framework, as \(S\to\infty\) the viscous penetration depth becomes vanishingly small and the gas in the gap has insufficient time to respond; consequently, the NSF prediction yields a diminishing force amplitude and a phase approaching \(-90^\circ\), corresponding to a purely inertia response.

In contrast, when rarefaction and finite molecular relaxation are accounted for, the high-frequency asymptotics are qualitatively different. In the kinetic description, once the oscillation period becomes comparable to or shorter than the collision relaxation time, the stress response is controlled by non-equilibrium momentum transfer rather than by viscous diffusion. As a result, GSIS predicts that the SFD magnitude tends to a finite constant as \(S\) increases, and the phase approaches \(0^\circ\), indicating a predominantly dissipative response. The same saturation trend is also observed for \(\delta_{rp}=1\) and \(10\), consistent with the fact that in the high-frequency limit the dominant contribution comes from collisionless (or weakly collisional) molecular interactions with the oscillating boundary, which do not vanish with increasing frequency. Similar qualitative behavior has been reported in simpler configurations; for example, Wu obtained an analogous high-frequency kinetic limit in the planar acoustic-wave problem~\cite{wu2016sound}.

\begin{figure}[t]
\centering
    \includegraphics[width=0.47\linewidth]{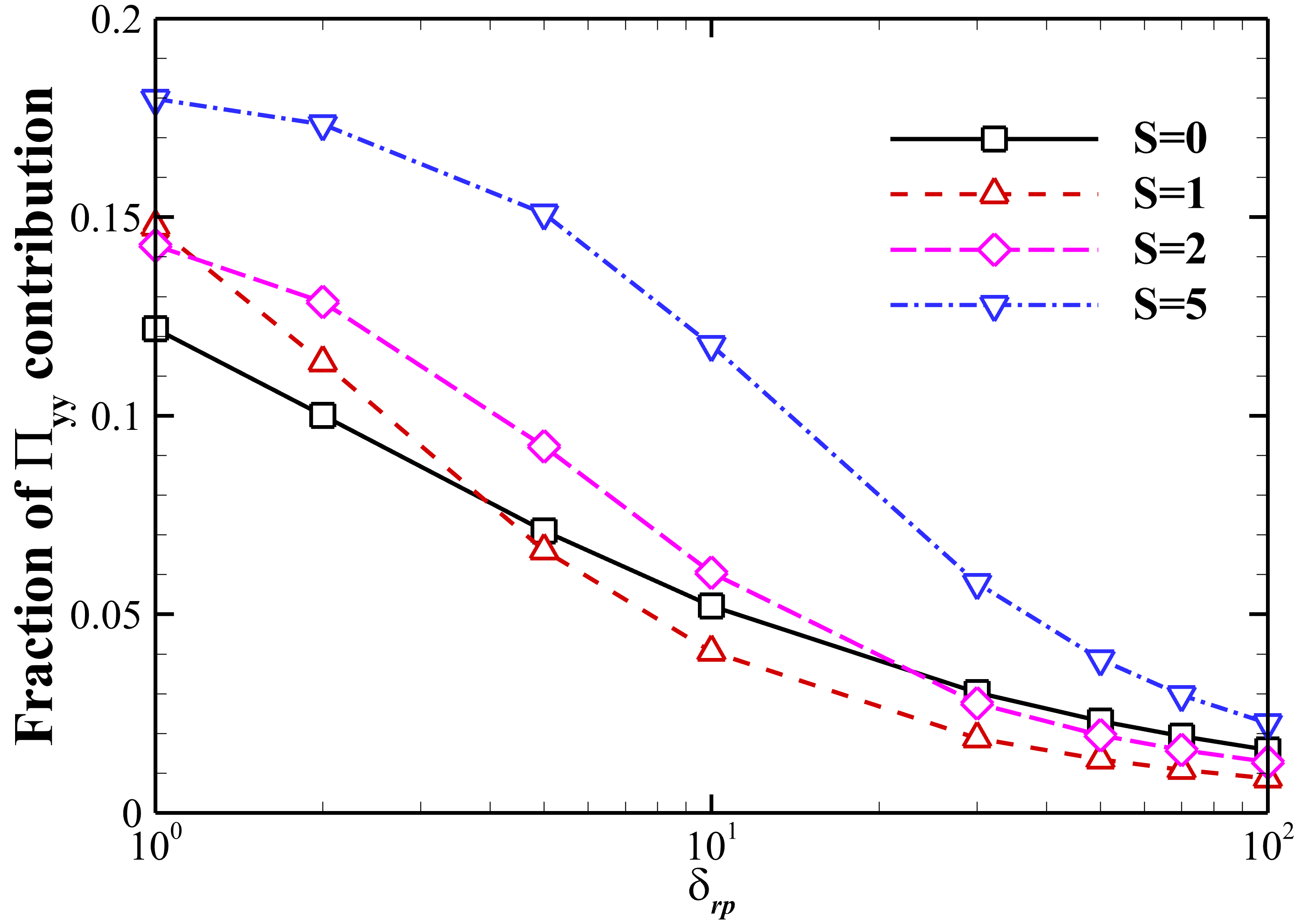}
    \hspace{0.2cm}
    \includegraphics[width=0.47\linewidth]{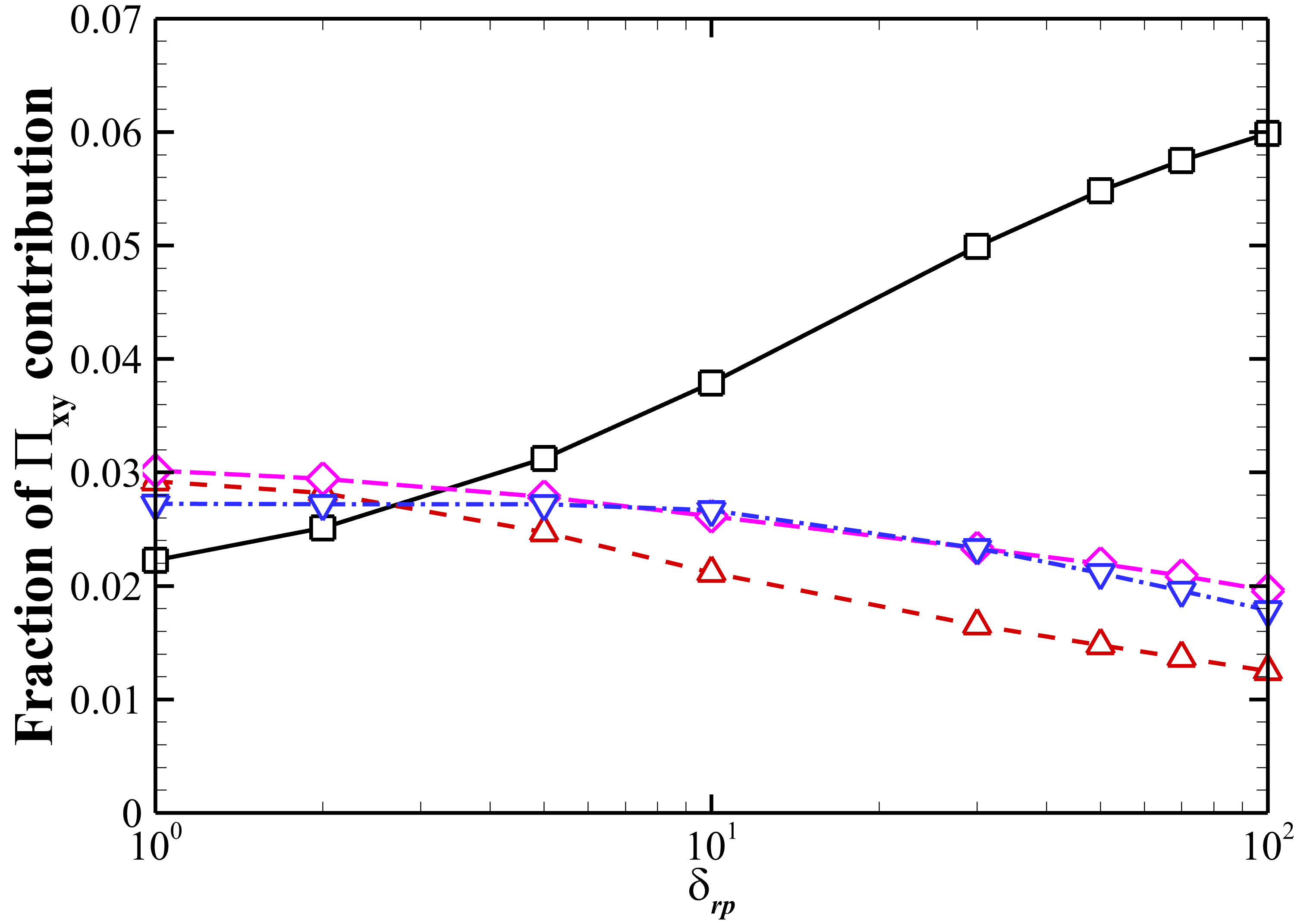}
    \caption{Magnitude fractions of the deviatoric-stress contributions to the SFD: (a) \( |F_{\Pi, yy}|/|F_{y}| \) and (b) \( |F_{\Pi, xy}|/|F_{y}| \) as functions of \(\delta_{rp}\) for several Strouhal numbers \(S\).}
    \label{fig:cantilever-fraction}
\end{figure}
To further assess the role of deviatoric stresses in the damping response, we quantify the respective contributions of \(\Pi_{yy}\) and \(\Pi_{xy}\) to \(F_y\). We define
\begin{equation}
F_{\Pi, yy}=-\int_{\Gamma_b}\Pi_{yy}\,dS_y,\quad
F_{\Pi, xy}=-\int_{\Gamma_b}\Pi_{xy}\,dS_x,
\end{equation}
and plot the magnitude fractions \( |F_{\Pi_{yy}}|/|F_y| \) and \( |F_{\Pi_{xy}}|/|F_y| \) as functions of \(\delta_{rp}\) for several Strouhal numbers \(S\) in Fig.~\ref{fig:cantilever-fraction}.

As shown in Fig.~\ref{fig:cantilever-fraction}(a), the \(\Pi_{yy}\) contribution increases as the flow becomes more rarefied (smaller \(\delta_{rp}\)) and as the oscillation frequency increases (larger \(S\)). This trend reflects the enhanced non-equilibrium response in rarefied and high-frequency conditions, where the deviatoric stress becomes more pronounced relative to the pressure contribution. In particular, at \(\delta_{rp}=1\), \(|F_{\Pi_{yy}}|/|F_y|\) exceeds \(10\%\) for the cases considered, indicating that neglecting \(\Pi_{yy}\) may lead to a noticeable error in SFD prediction in rarefied regimes.

In contrast, Fig.~\ref{fig:cantilever-fraction}(b) shows that the \(\Pi_{xy}\) contribution is comparatively small and only weakly dependent on \(\delta_{rp}\) and \(S\): it is typically around \(2\%\) and reaches about \(5\%\) only for the quasi-steady near-continuum case (e.g., \(\delta_{rp}=100\) and \(S=0\)). This behavior can be attributed to geometry: \(\Pi_{xy}dS_x\) mainly arises from the side faces associated with the cantilever thickness direction, and its contribution therefore scales with the thickness-related surface area. In typical MEMS configurations, where the thickness is much smaller than the width, this term is usually negligible; in the present setup, the thickness-to-width ratio is not extremely small, making the \(\Pi_{xy}\) fraction more visible, yet it remains secondary compared with the pressure and \(\Pi_{yy}\) contributions.

\section{Conclusions}\label{sec:conclusion}

In summary, we have developed a frequency-domain GSIS for the efficient simulation of oscillatory rarefied gas flows. By directly solving for the periodic steady response in the frequency domain, the proposed approach avoids long transient time-marching while retaining full kinetic fidelity across a wide range of spatial and temporal rarefaction levels. By reformulating the stress and heat-flux treatment in a symmetric and consistent manner, the frequency-domain GSIS has achieved robust and super-convergence across the investigated frequency range. In the near-continuum limit with small temporal Knudsen number, a Chapman–Enskog expansion has further indicated that the coupled kinetic–synthetic iteration is asymptotic-preserving, thereby enabling accurate solutions on coarse meshes.

We have demonstrated the numerical performance and accuracy in two challenging configurations: an oscillatory shear-driven flow between two eccentric cylinders and squeeze-film damping induced by an oscillating microcantilever above a substrate. In both cases, GSIS has attained convergence within 20–30 outer iterations over a broad parameter space, whereas CIS has required thousands of iterations in near-continuum, low-frequency conditions. The resulting reduction in time to solution has reached up to three orders of magnitude. 
Using the GSIS, we have discovered qualitatively distinct high-frequency cantilever dynamics predicted by kinetic theory, in contrast to continuum models. In particular, the results have revealed saturation of the damping magnitude and markedly different phase trends when the oscillation period becomes comparable to the molecular relaxation time.

\section*{Declaration of competing interest}
The authors declare that they have no known competing financial interests or personal relationships that could have appeared to
influence the work reported in this paper.


\appendix

\bibliographystyle{elsarticle-num}

\bibliography{cite}

\end{CJK}
\end{document}